\documentclass[]{aastex631}
\usepackage{xcolor}
\usepackage{soul}

\begin{document}
\title{Comparative study of the kinetic properties of proton and alpha beams   \\ in the Alfvénic wind observed by SWA-PAS onboard Solar Orbiter}

\correspondingauthor{Rossana DeMarco}
\email{rossana.demarco@inaf.it}

\author[0000-0002-0786-7307]{Roberto Bruno}
\altaffiliation{shared co-first authorship}
\affiliation{ INAF-Istituto di Astrofisica e Planetologia Spaziali, Via Fosso del Cavaliere 100, 00133 Roma, Italy.}

\author[0000-0002-7426-7379]{Rossana De Marco}
\altaffiliation{shared co-first authorship}
\affiliation{ INAF-Istituto di Astrofisica e Planetologia Spaziali, Via Fosso del Cavaliere 100, 00133 Roma, Italy.}


\author[0000-0003-2647-117X]{Raffaella D'Amicis}
\affiliation{ INAF-Istituto di Astrofisica e Planetologia Spaziali, Via Fosso del Cavaliere 100, 00133 Roma, Italy.}

\author[0000-0003-1059-4853]{Denise Perrone}
\affiliation{  ASI - Italian Space Agency, Rome, Italy.  }

\author[0000-0002-5002-6060]{Maria Federica Marcucci}
\affiliation{ INAF-Istituto di Astrofisica e Planetologia Spaziali, Via Fosso del Cavaliere 100, 00133 Roma, Italy.}
         
\author[0000-0002-6710-8142]{Daniele Telloni}    
\affiliation{ INAF - Osservatorio Astrofisico di Torino, Torino, Italy.}
\author[0000-0002-6433-7767]{Raffaele Marino}
\affiliation{ Univ Lyon, CNRS, \'Ecole Centrale de Lyon, INSA Lyon, Univ Claude Bernard Lyon I,
Laboratoire de M\'ecanique des Fluides et d'Acoustique, UMR5509, F-69134 \'Ecully, France.}
\author[0000-0002-5981-7758]{Luca Sorriso-Valvo}
\altaffiliation{Space and Plasma Physics, School of Electrical Engineering and Computer Science, KTH Royal Institute of Technology, Teknikringen 31, SE-11428 Stockholm, Sweden.}
\affiliation{   CNR, Istituto per la Scienza e la Tecnologia dei Plasmi, Via Amendola 122/D, 70126 Bari, Italy.}
\author[0000-0001-6970-8793]{Vito Fortunato}
\affiliation{Planetek Italia S.R.L., Via Massaua, 12, 70132 Bari, BA, Italy.}
\author{Gennaro Mele}
\affiliation{LEONARDO spa, Grottaglie, Taranto, 74023 Italy.}
\author{Francesco Monti}
\affiliation{TSD-Space, Via San Donato 23, 80126 Napoli, Italy.}
\author[0000-0002-9975-0148]{Andrei Fedorov}
\affiliation{ Institut de Recherche en Astrophysique et Planétologie, CNRS, Université de Toulouse, CNES, Toulouse, France.}
\author[0000-0003-2783-0808]{Philippe Louarn}
\affiliation{ Institut de Recherche en Astrophysique et Planétologie, CNRS, Université de Toulouse, CNES, Toulouse, France.}
\author[0000-0002-5982-4667]{Chris J. Owen}
\affiliation{ Mullard Space Science Laboratory, University College London, Holmbury St. Mary, Dorking, Surrey, RH5 6NT,UK.}
\author[0000-0002-4149-7311]{Stefano Livi}
\affiliation{ Southwest Research Institute, 6220 Culebra Road, San Antonio TX, 78238, USA.}




\begin{abstract}
The problems of heating and acceleration of solar wind particles are of significant and enduring interest in astrophysics. The interactions between waves and particles are crucial in determining the distributions of proton and alpha particles, resulting in non-Maxwellian characteristics including temperature anisotropies and particle beams. These processes can be better understood as long as the beam can be separated from the core for the two major components of the solar wind. We utilized an alternative numerical approach that leverages the clustering technique employed in Machine Learning to differentiate the primary populations within the velocity distribution, rather than employing the conventional bi-Maxwellian fitting method. Separation of the core and beam revealed new features for protons and alphas. We estimated that the total temperature of the two beams was slightly higher than that of their respective cores, and the temperature anisotropy for the cores and beams was larger than 1. We concluded that the temperature ratio between alphas and protons largely over 4 is due to the presence of a massive alpha beam, which is approximately 50\% of the alpha core. We provided evidence that the alpha core and beam populations are sensitive to Alfv\'enic fluctuations and the surfing effect found in the literature can be recovered only when considering the core and beam as a single population. Several similarities between proton and alpha beams would suggest a common and local generation mechanism not shared with the alpha core, which may not have necessarily been accelerated and heated locally.    
\end{abstract}



\section{Introduction}
\label{intro}


The solar wind is a turbulent quasi-collisionless magneto-fluid expanding in the interplanetary space at supersonic and super-Alfv\'enic speed. The velocity distributions of the ions forming the solar wind plasma generally exhibit conspicuous non-Maxwellian features \citep{asbridge1974,feldman1973a,feldman1973b,feldman1974,feldman1993,marsch1982a,marsch1982b,kasper2008,verscharen2011,marsch2012,matteini2007,matteini2013,goldstein1995,goldstein2000,goldstein2010} such as beams of accelerated ions and temperature anisotropies. 
\textcolor{black}{Temperature anisotropies can significantly impact particle dynamics and generate fluctuations. As stressed by \cite{matteini2013}, temperature anisotropy can provide a source of free energy for plasma instabilities. If the anisotropy exceeds a certain threshold, unstable fluctuations scatter particles towards a more isotropic structure. The dominant temperature direction loses energy, which gets transferred to other components and electromagnetic fluctuations. This leads to a decrease in kinetic energy and an increase in the local power of fluctuations, resulting in cooler particles.}
In particular, proton velocity distributions in fast Alfv\'enic streams are characterized by a largely anisotropic core, with $T_{\perp} > T_\|$, and by a secondary beam population.
On the other hand, velocity distributions within non-Alfv\'enic slow and intermediate wind speed often exhibit an isotropic core but an overall temperature anisotropy with $T_{\perp} < T_\|$, typically due to a high-energy tail or to the presence of a resolved proton beam
\citep{marsch1982b}.
Proton beams, identified for the first time by \cite{feldman1973a}, were studied in detail
using Helios observations
in the inner heliosphere, ACE and WIND/SWE at 1 au 
and Ulysses in the outer heliosphere and at high-latitude \citep[see the extensive review by][and references therein]{feldman1997}.
In Alfv\'enic high-speed streams, the proton beam abundance is usually around 10\% of the core proton population \citep{marsch1982b} while the drift velocity of the beam relative to the proton core tends to be quite larger than the Alfv\'en speed.

Alpha particles (i.e., fully ionized helium atoms, He$^{2+}$) represent the second most abundant ion population, accounting for $\sim 20\%$ of the solar wind mass density, which corresponds to $\sim 5\%$ of the total ion number density
\citep{neugebauer1962,neugebauer1966,asbridge1974,marsch1982a,yermolaev1997,kasper2007}. 
Alpha particle VDFs were studied in detail by \cite{marsch1982a} using Helios observations. Under Alfv\'enic wind conditions, this population exhibits a total temperature between $4$ and $5$ times that of the protons, well beyond an isothermal wind condition that can be recovered only within the slow non-Alfv\'enic wind \citep{kasper2008,maruca2013}. 
Moreover, this ion population is characterized by a temperature anisotropy with $T_\| > T_{\perp} $, opposite to that of the proton VDFs \citep{marsch1982b}, and by the sporadic presence of a well-resolved secondary beam \citep{feldman1973a,marsch1982a,nemecek2020}.
Similarly to the proton beam, alpha particles stream faster than the proton core population \citep{nemecek2020}. 
Within Alfv\'enic wind, this velocity drift is slightly less than the Alfv\'en speed \citep{marsch1982a,neugebauer1996}, while in the slow non-Alfv\'enic wind, proton and alpha populations often display no differential speed \citep{kasper2008,Maruca2012}. 

There is a complex interplay between plasma particles and electromagnetic waves that shapes the different ion distributions and regulates phenomena like beam generation, particle heating, and drift velocity \citep{marsch2018}. Ion-cyclotron wave absorption through cyclotron resonance \citep{isenberg2001,marschtu2001,hollweg2002,matteini2007,araneda2008,araneda2009,coleman1968,denskat1983,goldstein1994,leamon1998a,gary1999,he2011,he2015,jian2009}, stochastic heating of oblique kinetic Alfv\'en waves via Landau damping \citep{leamon1998b,leamon1999,leamon2000,howes2008} are the most likely mechanisms that produce non-Maxwellian features of the ion velocity distribution. 
Moreover, the energy necessary to activate these processes 
is provided by non-linear turbulent processes which, at fluid scales, transfer energy towards smaller and smaller scales to finally reach the kinetic regime \citep[and references therein]{tumarsch1995,brunocarbone2013,sorriso2019}.  
Thus, it is natural to expect that the above phenomena be more relevant within solar wind regions characterized by strong Alfv\'enic turbulence at fluid scales and only weakly collisional, like fast or slow Alfv\'enic streams since collisionality erodes non-Maxwellian features \citep{kasper2017}.
In the rest of the paper, we will compare for the first time the kinetic features of core and beam populations for both proton and alpha 
particles observed within an Alfv\'enic fast wind by \textcolor{black}{the Solar Wind Analizer (SWA) \citep{owen2020} onboard the Solar Orbiter mission \citep{muller2020}} at $0.58$ au from the Sun.  The possibility of separating these components of the particle velocity distribution is offered by 
the novel technique described in \cite{demarco2023}.  

\section{Method}
\label{metodo}
Clustering is used for grouping data based on a defined set of characteristics. It is one of the most widely used forms of unsupervised machine learning, which allows to reveal underlying patterns in the data. As opposed to the usual fitting procedure, our technique does not rely on approximating the distribution function with an ideal bi-Maxwellian, which would constrain de facto the physics embedded in the solar wind.  Instead, it tries to identify sub-populations within the overall data set, assigning to each observation a probability of belonging to a certain group. Then, the portion of the total velocity distribution function 
belonging to each ion family can be determined, even in the presence of overlapping. The entire process, from the data cleaning to the validation of the results, is described in \citet{demarco2023}, where it has been designed to identify up to three ion populations. For the present work, the algorithm has been upgraded to separate up to four ion populations.
The number of groups to identify is an input parameter in this process. The moments are then computed for each separated velocity distribution. 
\begin{figure*}
    \centering
    \includegraphics[width=18cm]{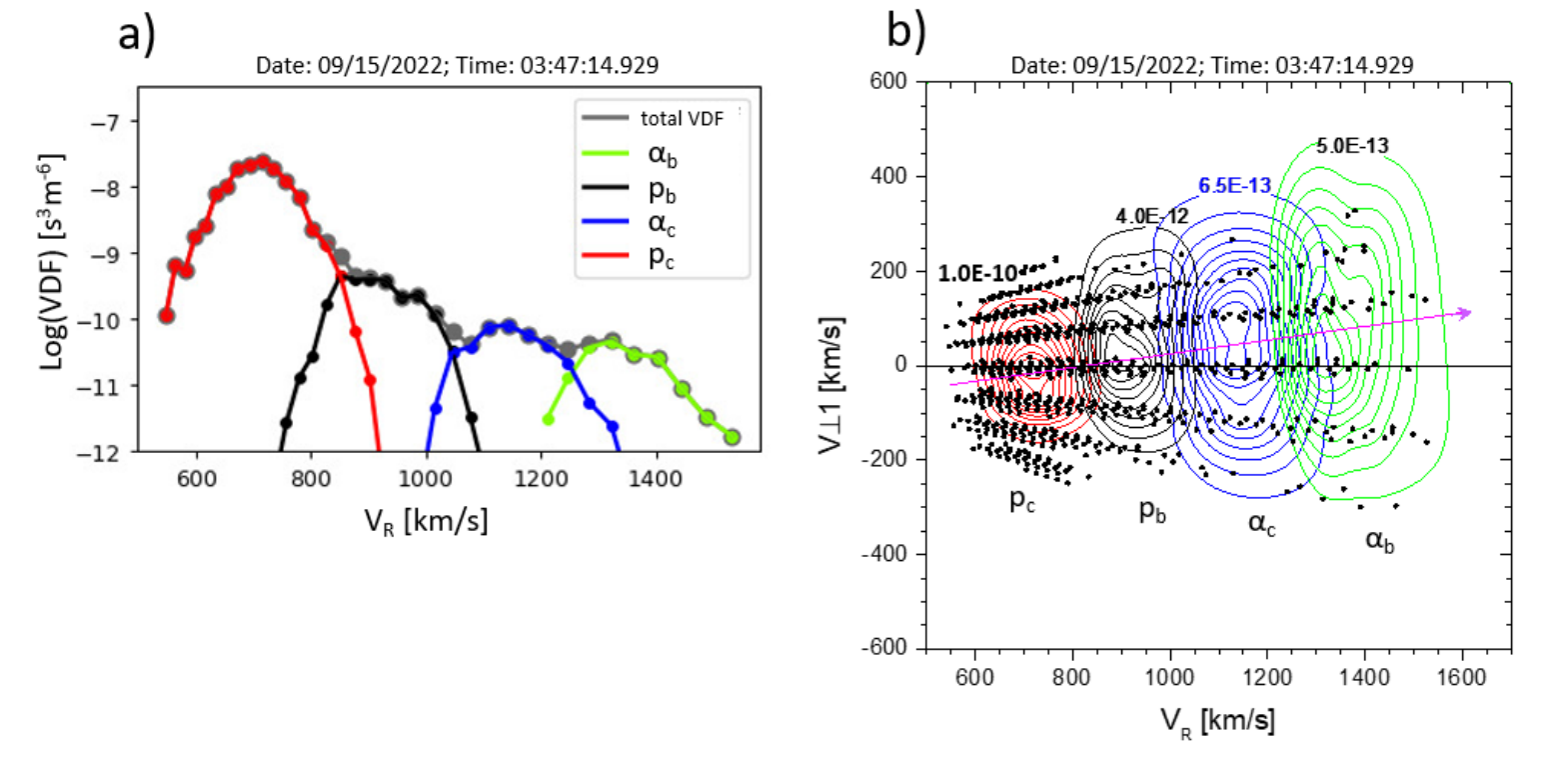}
    \caption{Panel a) The four main populations of a generic VDF are displayed in distinct colors. In particular, the VDF-1D versus radial velocity VR is integrated along the two perpendicular directions. Dots of different colors refer to different populations. In particular, dots in red, black, blue, and green refer to proton core, proton beam, alpha core, and alpha beam, respectively (see text for more details). PAS data are shown in the RTN reference system; panel b) contour lines of the same populations of panel a) whose VDF has been integrated along the direction perpendicular to the V-B plane shown in this panel. The outermost contour refers to VDF values that are $10\%$ of the innermost contour. The pink arrow indicates the local magnetic field.}
    \label{Fig01}
\end{figure*}

Before we describe the analysis results, we provide an example of our code's capabilities as shown in Figure \ref{Fig01}.
Panel a) shows the VDF-1D versus radial velocity VR integrated along the two perpendicular directions. Dots in red, black, blue, and green refer to proton core, proton beam, alpha core, and alpha beam, respectively. The gray dots, mostly covered, refer to the whole VDF. The \textcolor{black}{Proton and Alpha Sensor (PAS), which is one of the SWA plasma instruments}, \citep[][see section \ref{analisi} ]{owen2020} does not have a time-of-flight section able to discriminate among different ions but simply performs an energy-over-charge $(E/q)$ selection. Consequently, the alpha particles appear to move faster than protons by a factor $\sqrt{2}$. Panel b) shows contour lines for the same four populations rotated and integrated in the direction perpendicular to the $B-V$ plane. This plane contains the radial direction R and the local magnetic field vector \textcolor{black}{$\vec{B}$},
indicated by the pink arrow. This plot already provides some interesting information that characterizes the  different populations. In particular, the contours show that the temperature increases moving from the proton core towards the alpha beam and that the four populations are quite aligned with the direction of the local magnetic field. These aspects will be discussed more in-depth later in the paper.
Our code is sometimes unable to accurately identify sub-groups in the distribution (see \citet{demarco2023}). In these cases, the distribution was removed from the data set. 
For the data interval analyzed here, described in the next section, we discarded $10.5\%$ of the data, leaving $55931$ VDFs for the in-depth analysis.

\section{Data Analysis}    
\label{analisi}
The data used in this paper are provided by the Solar Wind Analyser (SWA) \citep{owen2020} and by the magnetometer (MAG) \citep{horbury2020} on board the Solar Orbiter mission \citep{muller2020}. SWA is a plasma suite consisting of three sensors: the Electron Analyzer Sensor (EAS), the hot ion sensor (HIS), and the proton alpha sensor (PAS). In particular, this study focuses on the 3D velocity distribution functions (VDFs) of protons and alpha particles measured by PAS with a cadence of 4 s and sampling time of $\sim 1$ s for each single distribution. 
Such a rapid sampling time protects us from possible instrumental effects in determining the kinetic characteristics of the plasma  \cite[see papers by][]{verscharen2011,perrone2014,nicolaou2019,demarco2020}.

As we will see later in the paper, we chose to analyze the high-speed region of an Alfv\'enic high-speed stream observed by Solar Orbiter on September 2022 at a distance of 0.58 au from the Sun, just before the second perihelion of the nominal phase. 
We used PAS 3D VDFs and ground moments (i.e. number density, velocity vector, and temperature computed from the pressure tensor) derived from the VDFs at 4~s cadence. Magnetic field measurements, provided by MAG, are averaged at the plasma sampling time. Data are available on the Solar Orbiter Archive (SOAR) (http://soar.esac.esa.int/).



\begin{figure}
    \centering
    \includegraphics[width=12cm]{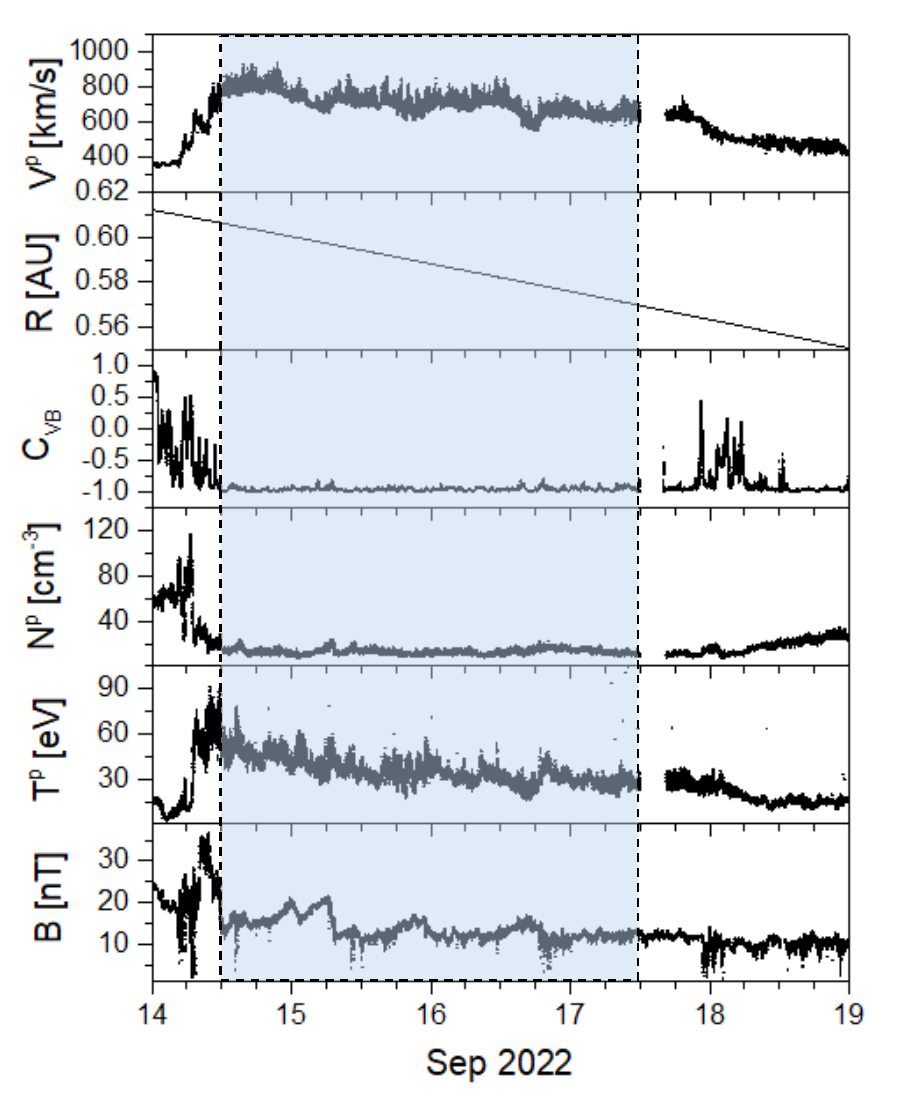}
    \caption{Time series of relevant parameters of the fast wind stream observed by Solar Orbiter in mid-September 2022. Plasma data shown here are the L2 ground moments downloaded from the SOAR archive. From top to bottom: Proton bulk speed, $V_p$; heliocentric distance, R; v-b correlation coefficient computed over a 30 min sliding window, $C_{VB}$; proton number density, $N_p$; proton temperature, $T_p$; magnetic field magnitude, $B$.}
    \label{Fig02}
\end{figure}

Figure \ref{Fig02} shows the time series of relevant parameters: the proton bulk speed, $V_p$; the heliocentric distance, R; the velocity-magnetic field (v-b) correlation coefficient computed over a 30~min sliding window, $C_{VB}$; the proton number density, $N_p$; the proton temperature, $T_p$ and the magnetic field magnitude, $B$. Plasma data, in this case, are the 4~s L2 ground moments from SOAR. These moments refer only to protons and the elimination of alpha particles was achieved by cutting the 1-D distribution at the saddle point between protons and alpha particles. 
This typical fast stream is characterized by three main regions: the compressive interface region, the high-speed plateau, and the rarefaction region. $C_{VB}$ is computed using a typical Alfvénic scale \citep[e.g.][]{marschtu1990} and it is used as a quick look parameter to identify Alfvénic fluctuations. Our focus will be on the Alfvénic part of the stream, which is identified by the light blue box and covers most of the high-speed plateau. This region, starting on the $14^{th}$ of September 2022 at 14h:24m:03s and ending on the $17^{th}$ at 11h:53m:07s, is characterized by $C_{VB}$ close to 1 and typically has one-sided, large amplitude velocity fluctuations, as described by \cite{matteini2015}. 


\subsection{Identification of ion populations}
\begin{figure}
    \centering
    \includegraphics[width=12cm]{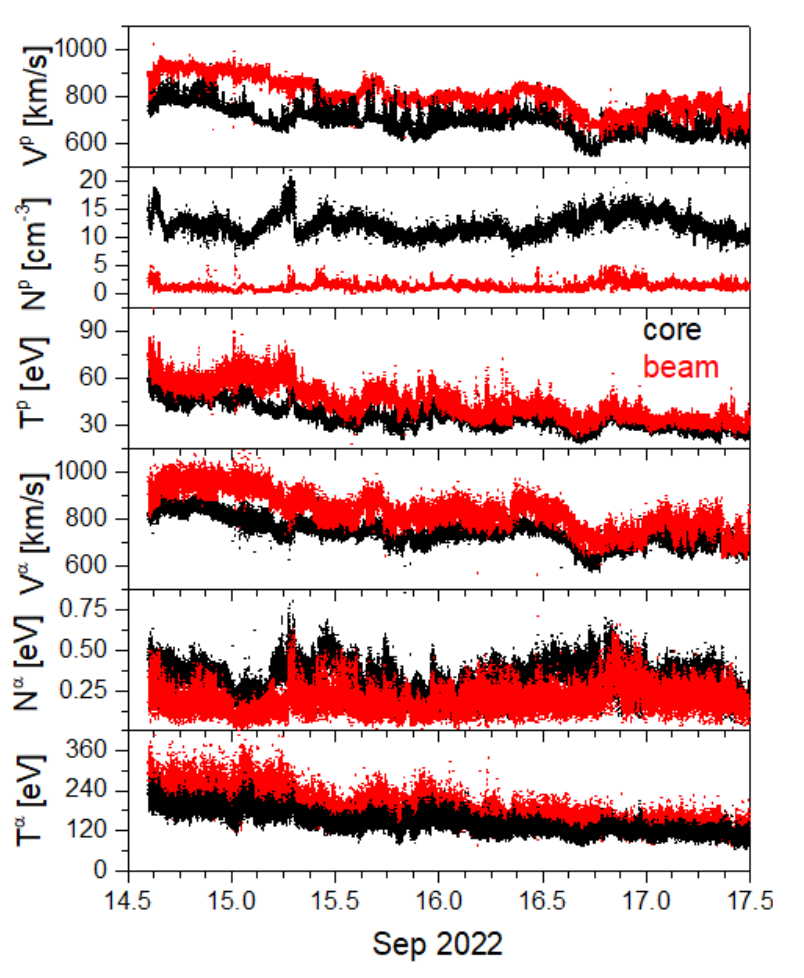}
    \caption{Time series of plasma parameters of the different populations derived from the clustering code, applied to the interval identified by the light blue box in Fig \ref{Fig02}. From top to bottom: proton bulk speed, $V_p$, number density, $N_p$ and total temperature, $T_p$ (black and red for proton core and proton beam, respectively); alpha particles bulk speed, $V_{\alpha}$, number density, $N_\alpha$ and total temperature, $T_{\alpha}$ (black and red for alpha core and alpha beam, respectively).}
    \label{Fig03}
\end{figure}

We applied our clustering code to 
identify the four ion populations in
the PAS VDFs during the time interval of interest. In Figure \ref{Fig03}, we show the velocity, number density, and total temperature time series for the core and beam of both protons and alphas.
The behavior of proton and alpha beams is pretty similar. Generally, beams are faster, less dense, and hotter than the cores. However, there is a significant difference in their relative number density. The alpha beam has more relevance compared to the alpha core, whereas the proton beam is much less relevant in this regard (see later in the paper for details). 

The clustering technique allows us to analyze VDFs by choosing either 2 or 4 populations identification, depending on how many subgroups we want to identify. We will compare the results of our 4 population analysis with those of our 2 population analysis and with previous literature that did not distinguish between the beam and the core.

Starting from the distribution of the relative number density of the total alpha population $N_\alpha$ (core+beam) with respect to the total proton population $N_p$ (core+beam), not shown here, we found that it is quite symmetric with respect to its average value of $0.039\pm0.007$, very much in agreement with previous determinations in the fast solar wind \citep[][among others]{neugebauer1962,neugebauer1966,asbridge1974,yermolaev1997,kasper2007}. 
%
\begin{figure*}
\centering
\includegraphics[width=18cm]{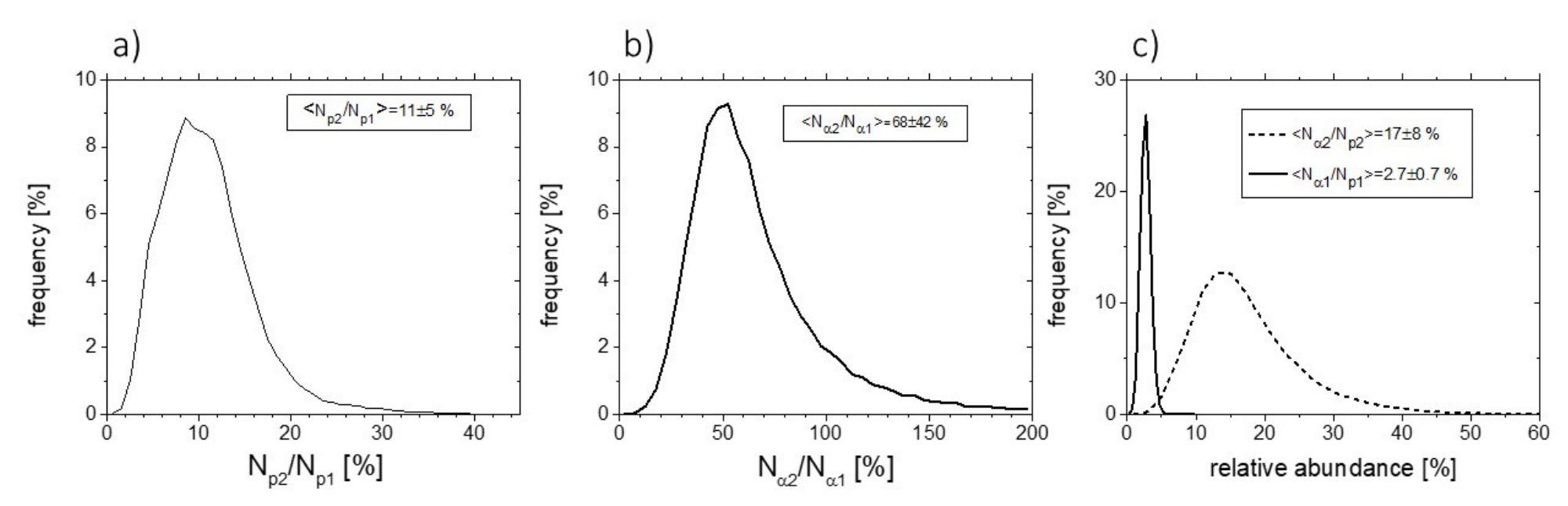}
\caption{Panel a): histogram of the relative density between proton beam and core; panel b): histogram of the relative density between the alpha beam and core; panel c): relative alpha abundance with respect to protons for the core and the beam
}
\label{Fig04}
\end{figure*}
Panel a) of Figure \ref{Fig04} shows the distribution of the relative number density of the proton beam $N_{p2}$ with respect to the proton core $N_{p1}$. The distribution is peaked at slightly less than $10\%$ and has an average value of $11\pm5\%$. In addition, it has a tail that stretches towards higher values. Proton beams around $10\%$ of the proton core are in good agreement with previous estimates reported in literature \citep{marsch1982b}. Less common are determinations of the relative density of the alpha beam $N_{\alpha 2}$ with respect to the alpha core $N_{\alpha 1}$ although previous estimates have already been given by  \cite{asbridge1974}.
Our analysis shows in panel b) of Figure \ref{Fig04} that this ratio peaks around $50\%$ and has a heavy tail towards higher values which strongly contributes to setting the average value at $68\pm42\%$. Although this last value probably might suffer inaccurate determinations of the alpha density because of low counts in the VDF, the most probable value shows that the beam for the alpha population is much more relevant than in the case of protons. Panel c), shows the relative helium to hydrogen abundance for the core and the beam. The alpha core population represents about $3\%$ of the proton core population while this ratio increases noticeably towards $17\%$ for the beam. Our findings partially confirm those of \cite{asbridge1974}. However, differences in final values may result from factors such as radial distance, wind speed, and Alfv\'enicity.  







\begin{figure*}
    \centering
   \includegraphics[width=18cm]{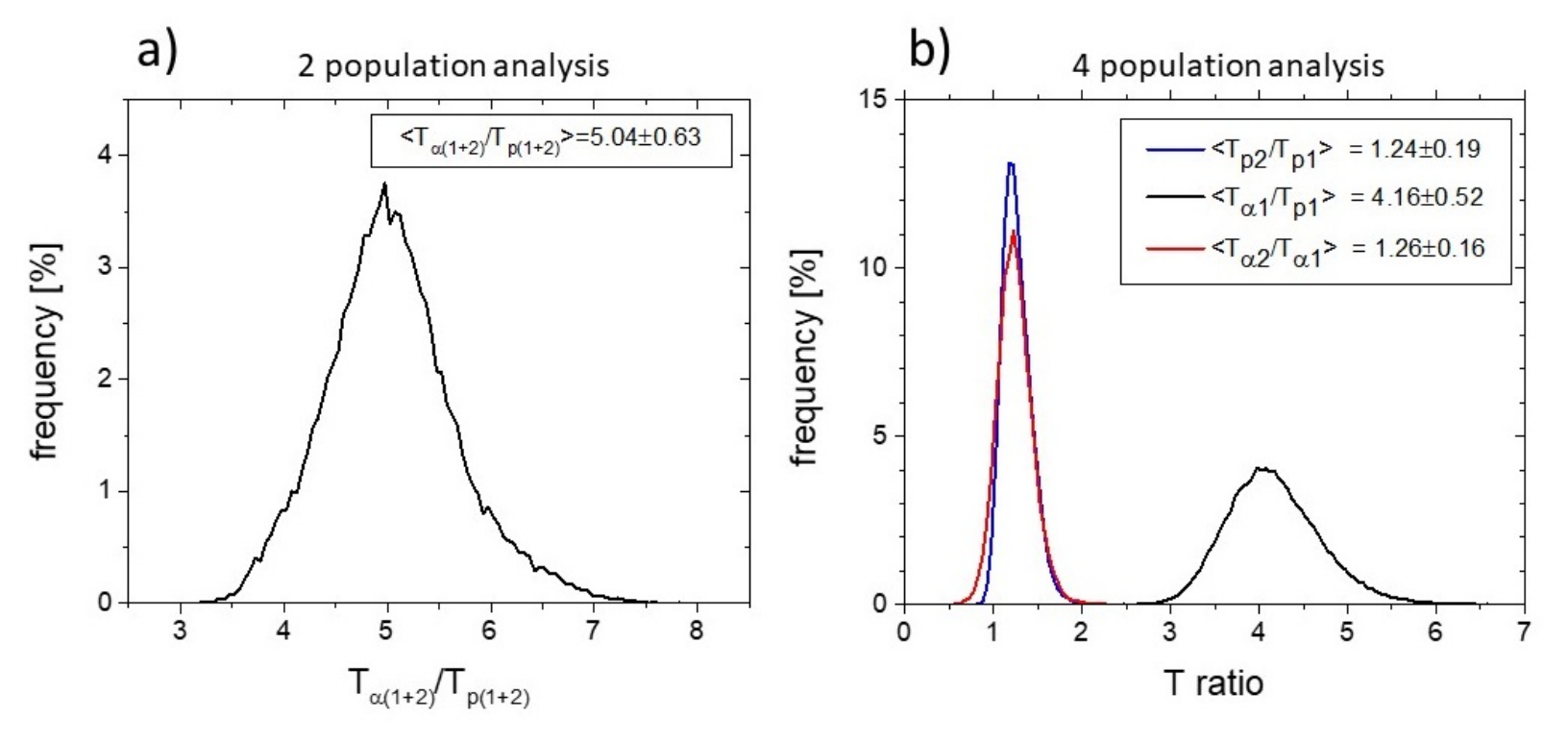}
   \caption{Panel a) Histogram of total temperature ratio between alpha and proton populations from the 2 populations analysis. Panel b) black line: total temperature ratio between alpha core and proton core; red line: total temperature ratio between the alpha beam and alpha core; blue line: total temperature ratio between proton beam and proton core from the 4 populations analysis.}
    \label{Fig05}
\end{figure*}

\subsection{Temperature and temperature anisotropy}
In Figure \ref{Fig05}, panel a) shows the histogram of the total temperature ratio $T_{\alpha(1+2)}/T_{p(1+2)}$ between alpha and proton populations based on the two-population analysis. \textcolor{black}{The subscripts $\alpha(1+2)$ and $p(1+2)$ indicate that core and beam are treated as a single population.} The histogram is quite symmetric around its peak whose value is remarkably close to the average value of the distribution around $5$. This result confirms previous analyses based on ACE observations \citep{kasper2008,maruca2013} that showed a robust tendency of this ratio towards a value between $4$ and $5$ with a remarkable fraction over $5$ within the high-speed wind. Large values of this ratio up to $5$ or $6$ had already been observed in previous investigations \citep{feldman1974,neugebauer1976, feynman1975,kasper2017}.
Clearly, the fast wind is not an isothermal fluid since in this case, the histogram would peak around $1$ as found by \cite{kasper2008,maruca2013} within the slow and more collisional wind. The anomalous value of $4$ for this ratio is expected for an equal thermal velocity for protons and alphas, probably due to heating processes active during the wind expansion which preferentially heat the minority species \citep{tumarsch2001,marschtu2001}. Moreover, values of $T_{\alpha(1+2)}/T_{p(1+2)} \ge 5$ require an additional anomalous heating mechanism as invoked by \cite{kasper2008}. 
On the other hand, distributions derived from our 4-population  analysis indicate that we can probably relax about this additional anomalous heating mechanism. Panel b) shows three histograms for the following ratios: $T_{p2}/T_{p1}$  (blue line); $T_{\alpha 2}/T_{\alpha 1}$ (red line); $T_{\alpha 1}/T_{p1}$ (black line), where $T_{p2}, T_{p1}, T_{\alpha 2}, T_{\alpha 1}$ are the total temperature of proton beam, proton core, alpha beam and alpha core, respectively. The histogram of $T_{\alpha 1}/T_{p1}$ clearly peaks around $4$ and values larger than $5$ represent only about $5\%$ of the whole distribution. The other two histograms are quite similar and peak at values slightly larger than $1$ indicating that beams are slightly hotter than the respective cores and probably share a common generation mechanism. Thus, it appears that the anomalous increase in temperature for the alpha particles can be attributed to the presence of the rather massive alpha beam drifting along the local magnetic field (refer to Figure \ref{Fig04}). This beam would cause an increase in the parallel temperature, and as a result, an increase in the overall alpha temperature, if we consider the alpha core and beam as a single population.


\begin{figure*}
    \centering
    \includegraphics[width=18cm]{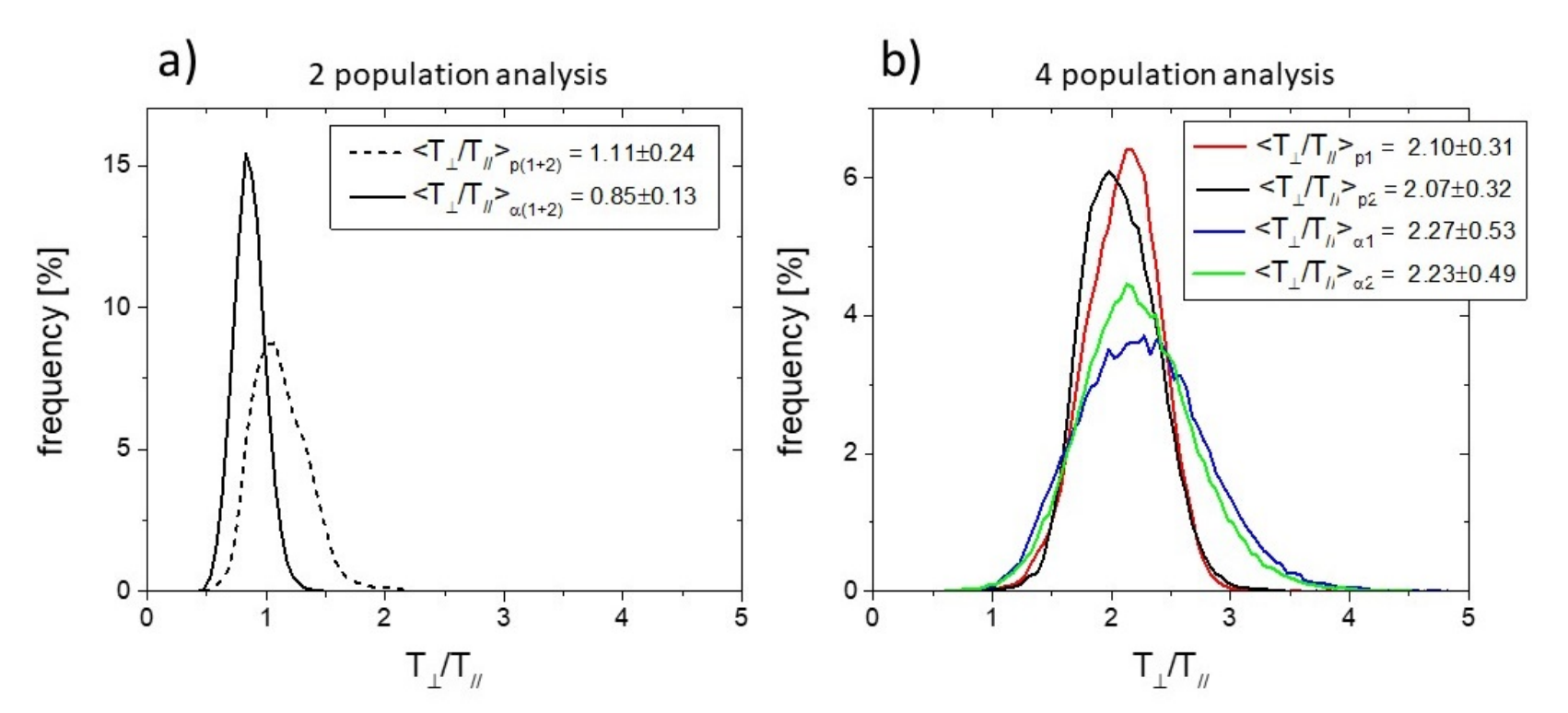}
   \caption{Panel a): Histograms of temperature anisotropy for protons (dashed line) and alphas (solid line) from the 2 populations analysis. Panel b) Histograms of temperature anisotropy for proton core (red line), proton beam (black line), alpha core (blue line), and alpha beam (green line) from the 4 populations analysis.}
    \label{Fig06}
\end{figure*}


The temperature anisotropies for protons and alphas are displayed in Figure \ref{Fig06}. Panel a) presents results from the analysis of 2 populations, \textcolor{black}{and, as in the case of the previous Figure, the subscripts $\alpha(1 + 2)$ and $p(1 + 2)$ indicate that core and beam are treated as a single population.} The graph shows that the alphas have a slightly lower anisotropy than 1, while the protons are slightly above 1. These findings are consistent with previous observations made by \cite{marsch1982a,marsch1982b} based on Helios data.  
In contrast, panel b) shows that the temperature anisotropy for all 4 populations is approximately 2. The significant difference between the results of the two analyses is due to the presence of the beam for both alphas and protons, which increases the parallel temperature, especially for the alphas because the beam is a good fraction of the core. 
Temperature anisotropy larger than $1$ of the proton core within fast wind streams in the inner heliosphere has been largely reported by  \cite{marsch1982b} while similar observations for the proton beam, in the outer heliosphere, have been reported by \cite{goldstein2010} who found that for large drift speed, as in our case, $T_\perp /T_{\|} > 1$. On the other hand, the general trend for observations in the outer heliosphere \citep{matteini2013}, would indicate
an anisotropy $> 1$ for the core distribution and $< 1$ for the beam distribution.
Helium ions temperature anisotropy was largely studied by \cite{marsch1982a} who concluded that 
in high-speed streams there was a slight indication for $T_\perp /T_{\|} < 1$ for the core part of the distribution. However, \cite{marsch1982a} did not separate the beam from the core and it is not clear how much magnetic field-aligned temperature bulges, reported by the authors, influenced their determinations. Thus, to our knowledge, our results provide the first determination of the temperature anisotropy of the alpha beam.

\subsection{Velocity drift}
\begin{figure*}
    \centering
    \includegraphics[width=18cm]{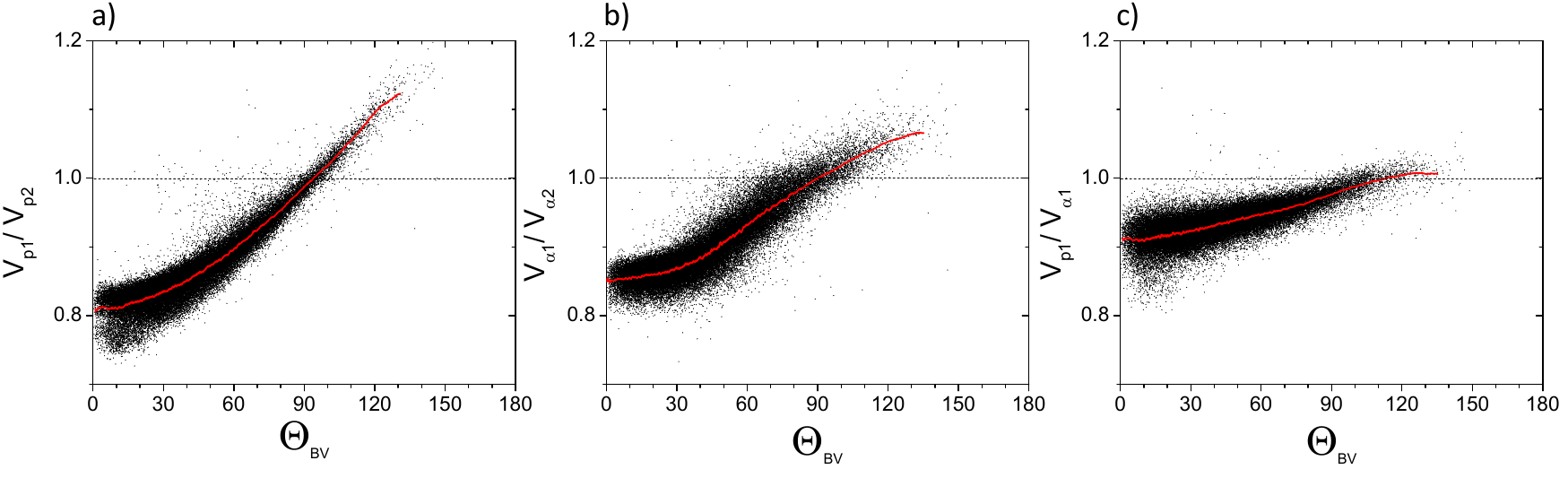}
   \caption{The three panels show distributions of various velocity ratios versus the angle $\Theta_{BV}$. 
   Panel a) the ratio between proton core and proton beam speeds; panel b) the ratio between alpha core and alpha beam speeds; panel c) the ratio between proton core and alpha core speeds. The red line represents a 500 points moving average drawn to guide the eye.}
    \label{Fig07}
\end{figure*}

\begin{figure*}
    \centering
    \includegraphics[width=18cm]{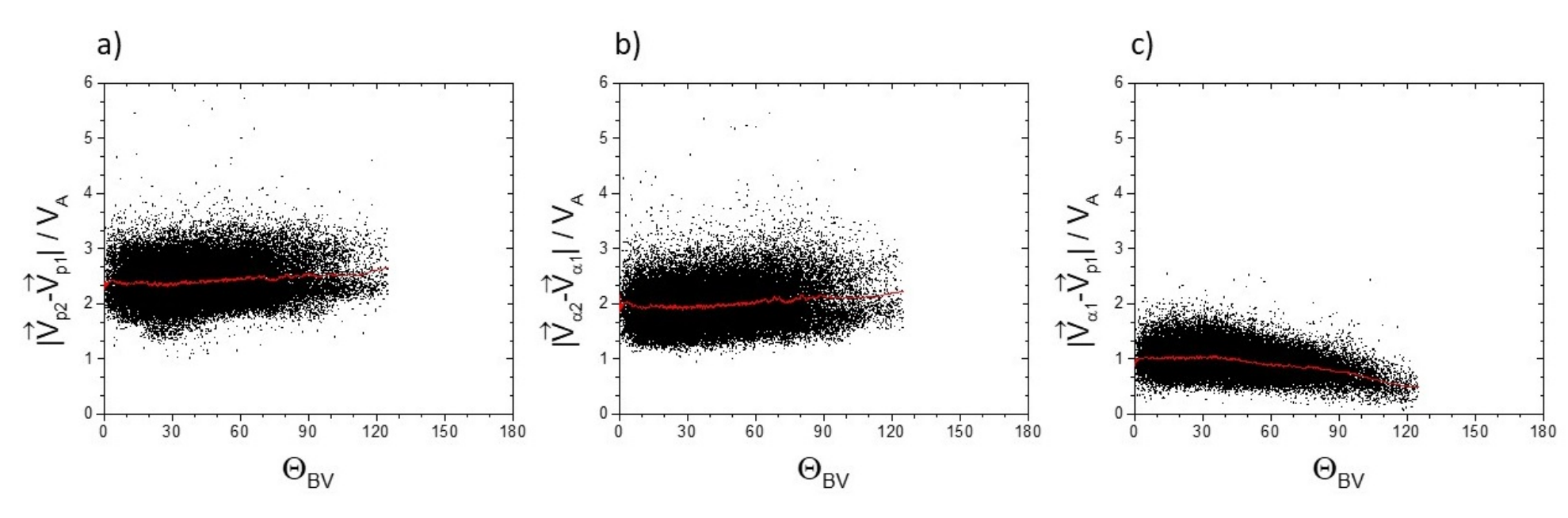}
   \caption{The three panels show the distributions of various velocity drifts, normalized to the local Alfv\'en velocity, versus the angle $\Theta_{BV}$. 
   Panel a)  the absolute value of the vector difference between proton beam and proton core velocities;
   panel b)  the absolute value of the vector difference between alpha beam and alpha core velocities;
   panel c)  the absolute value of the vector difference between alpha core and proton core velocities. The red line represents a 500 points moving average drawn to guide the eye.
   }
    \label{Fig08}
\end{figure*}

Following \cite{berger2011}, \cite{matteini2013} and \cite{durovcova2019}, we show in Figure~\ref{Fig07}   \textcolor{black}{distributions}
of velocity ratios versus the angle $\Theta_{BV}$, i.e. the angle formed by the proton core velocity vector and the local magnetic field vector. Symbols $V_{p1}, V_{p2}, V_{\alpha 1}$ and $V_{\alpha 2}$ refer to proton core, proton beam, alpha core, and alpha beam velocities, respectively. \textcolor{black}{The red line is a 500 points moving average drawn just to guide the eye.}
The local magnetic field vector $\vec{B}$ is the average vector obtained from 8 Hz magnetic field data averaged within PAS sampling time ($\sim$  1 s). Due to the fact that minor populations drift with respect to the proton core along the local magnetic field direction, we expect to see a dependence of this parameter on the angle $\Theta_{BV}$. 
Panel a) and c) are qualitatively similar to those reported by \cite{durovcova2019} for fast wind.  
We limit ourselves to a qualitative comparison since our analysis was performed for a much shorter time interval, selected at a fixed heliocentric distance, and, moreover, \cite{durovcova2019} do not separate the core and beam of the alpha population. The novelty with respect to \cite{durovcova2019} is therefore presented in panel b) where we show the ratio between alpha core and alpha beam speed. Comparing panel a) and panel b) we notice that the distribution for the alphas is less steepened than that for the protons, starts around $0^\circ$ at higher values, and tends to flatten out around $90^\circ$.  
\textcolor{black}{This does not mean that the drift velocity of each sub-population disappears for $\Theta_{BV}\rightarrow 90^\circ$, but only that the total velocity, which results from the vector sum of the proton core velocity plus the drift velocity of the sub-population, tends to be equal to the proton core velocity. The drift velocity of each sub-population, which is estimated as the absolute value of the vector difference between the velocity of the proton core and the velocity of the sub-population, normalized to the local Alfv\'en speed, remains relatively unaffected by the orientation of the magnetic field, as shown in Figure \ref{Fig08}, especially for the two beams (panels a and b). On the contrary, panel c shows the alpha core drift velocity slightly decreasing for angles larger than $40^\circ$ after a plateau around 1. 
}

\begin{figure}
    \centering
    \includegraphics[width=12cm]{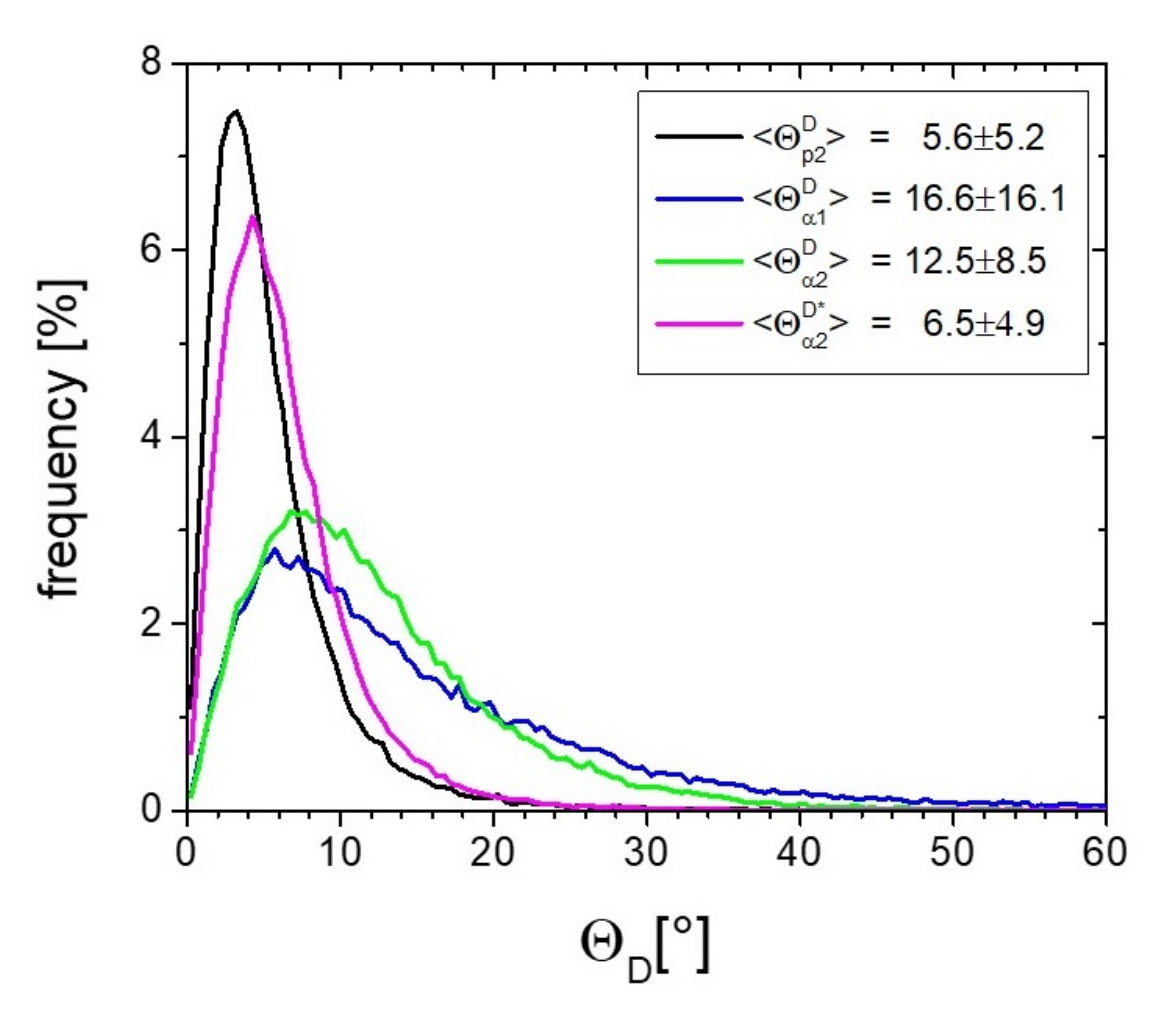}
   \caption{Histograms of the drift angle for the proton beam (black line) and alpha core (blue line) with respect to the proton core, and drift angle of alpha beam (green line) with respect to the alpha core.  This angle is measured between the local magnetic field and the drift velocity direction. The drift angle of the alpha beam with respect to the proton core is also shown by the magenta line.}
    \label{Fig09}
\end{figure}

Distributions of the drift angle for the proton beam  ${\Theta_{p2}^{D}}$ (black line), alpha core ${\Theta_{\alpha_1}^{D}}$ (blue line) and alpha beam ${\Theta_{\alpha_2}^{D}}$ (green line) are shown in Figure \ref{Fig09}. \textcolor{black}{This angle is measured between the local magnetic field orientation and the drift velocity direction of each sub-population. }
The drift velocity vectors for the proton beam and alpha core are estimated with respect to the proton core \textcolor{black}{velocity}. The alpha beam velocity drift vector is estimated with respect to the alpha core \textcolor{black}{velocity}. The figure also includes the average value for each distribution. These three populations are all quite aligned to the local magnetic field direction in agreement with the previous findings reported in literature \citep{marsch2018,marsch1982a,marsch1982b}.
However, the proton beam is more aligned to the local field than the alpha core and \textcolor{black}{the alpha} beam. The peak of the distribution is around $3^\circ$ with an average value of $5.6^\circ$. Moreover, the distribution is quite focused around this value having a standard deviation $\sigma=\pm 5.2^\circ $. On the contrary, alpha core and beam distributions are much less focused being characterized by a long tail. The peak of the core distribution is $\sim 6^\circ$ with an average value of $16.6^\circ \pm16.1^\circ$ while the peak for the beam is $\sim 7^\circ$ with an average value of $12.5^\circ\pm8.5^\circ$. 

\textcolor{black}{According to \cite{zhu2023a}, linear Vlasov theory posits that the presence of fast magnetosonic and/or whistler waves induces fluctuations in the velocities of both the proton core and beam, consequently disrupting the alignment of their velocity drift with the magnetic field. Subsequent analysis by the same authors, leveraging Solar Orbiter observations, revealed distinct time intervals characterized by the prevalence of circularly polarized fast magnetosonic/whistler (FM/W) waves. Within these intervals, the proton beam exhibited clear non-alignment with the magnetic field. In particular, their investigation unveiled that the non-aligned, fluctuating velocities within the beam population play a pivotal role in fostering the amplification of these waves. The meticulous analysis performed by these authors revealed that the peak of the probability density function of the angle between the proton beam drift velocity and the local field was a few degrees, remarkably in agreement with our findings.
This observation is interesting because, despite the $5^\circ$ angular resolution of PAS, we can discern plasma parameters with an uncertainty of under $5^\circ$, as remarked by \cite{zhu2023a}. }
It would be intriguing to explore whether these fluctuations have a comparable impact on the misalignment between the alpha core and alpha beam velocity drifts.
\textcolor{black}{In this context, caution is in order when dealing with the alpha distribution. The fact that alphas exhibit significantly lower number density and are detected at twice the energy of protons \citep{owen2020}, leading to a less distinct peak in the distribution, prompts further investigation. This entails disentangling potential instrumental effects, which may influence the angular distributions depicted in Figure~\ref{Fig09}, from underlying physical phenomena.
Finally, it is worth considering that the reason for the increased uncertainties could be because the drift velocities of alpha beam with respect to alpha core and those of alpha core with respect to proton core are usually smaller than the drift of proton core with respect to proton beam, and any displacement perpendicular to the magnetic field would result in a greater drift angle compared to a similar displacement on proton beam-proton core drift velocity. }

\begin{figure}
    \centering
    \includegraphics[width=17cm]{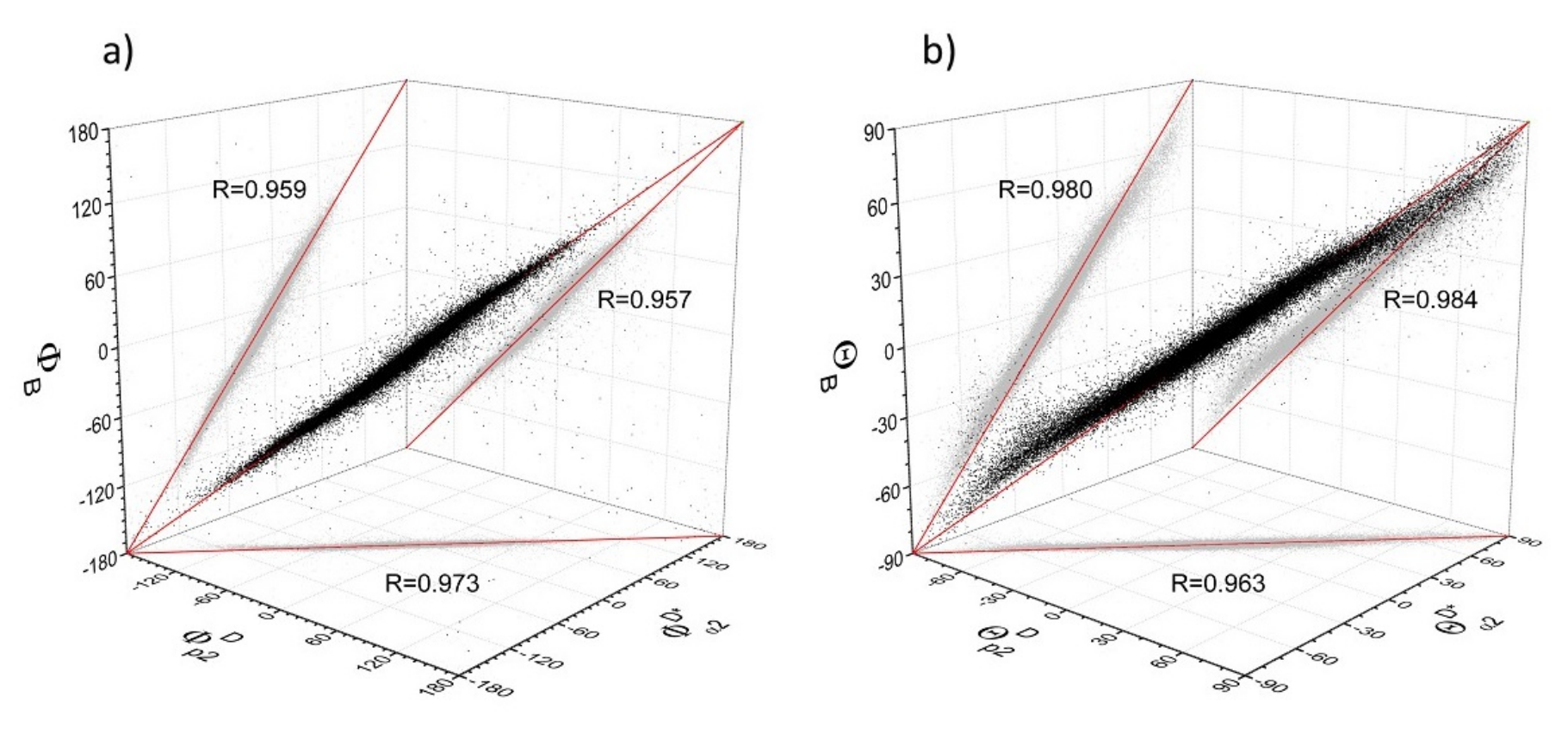}
   \caption{Panel a): 3D distribution of the azimuthal angles, in RTN reference system, of the magnetic field ($\Phi_B$), the proton beam drift velocity ($\Phi^D_{p2}$) and the alpha beam drift velocity with respect to the proton core ($\Phi^{D*}_{\alpha 2}$);  Panel b): 3D distribution of the polar angles, in RTN reference system, of the magnetic field ($\Theta_B$), the proton beam drift velocity ($\Theta^D_{p2}$) and the alpha beam drift velocity with respect to the proton core ($\Theta^{D*}_{\alpha 2}$). Diagonal red lines are marked to guide the eye. The values of the linear correlation coefficient for each pair of variables are displayed.}
    \label{Fig10}
\end{figure}
\begin{figure}
    \centering
    \includegraphics[width=17cm]{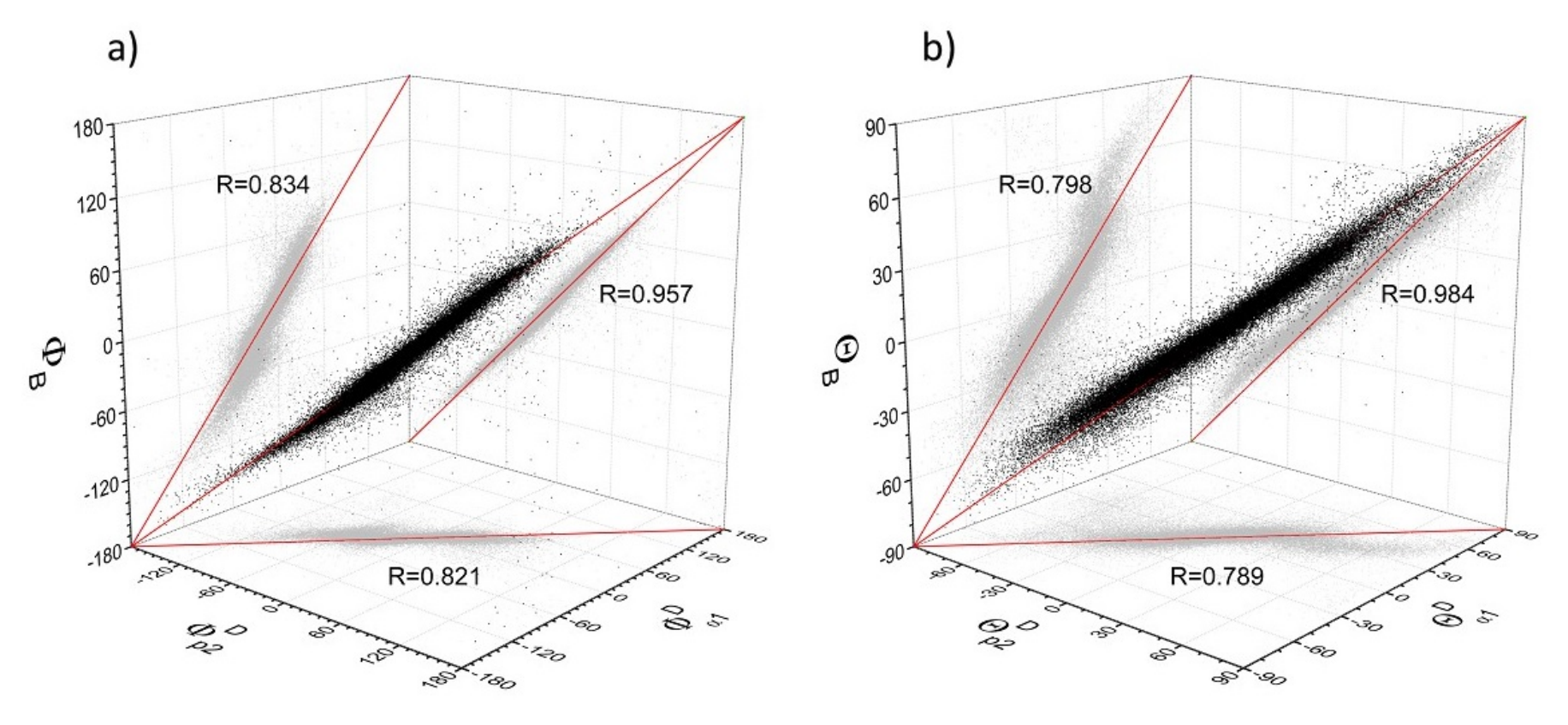}
   \caption{Panel a): 3D distribution of the azimuthal angles, in RTN reference system, of the magnetic field ($\Phi_B$), the proton beam drift velocity ($\Phi^D_{p2}$) and the alpha core drift velocity ($\Phi^D_{\alpha 1}$); Panel b): 3D distribution of the polar angles, in RTN reference system, of the magnetic field ($\Theta_B$), the proton beam drift velocity ($\Theta^D_{p2}$) and the alpha core drift velocity ($\Theta^{D}_{\alpha 1}$);  Diagonal red lines are marked to guide the eye. The values of the linear correlation coefficient for each pair of variables are displayed.}
    \label{Fig11}
\end{figure}
Interestingly, Figure \ref{Fig09} shows also the distribution of the drift angle of the alpha beam \textcolor{black}{velocity drift} with respect to the proton core, as indicated by the magenta line. There is a remarkable similarity with the proton beam distribution being characterized by an average value of $6.5^\circ\pm4.9^\circ$.  \textcolor{black}{It is currently unclear why the drift velocity of the alpha beam is more aligned with the local magnetic field when estimated with respect to the proton core rather than the alpha core. This finding is unexpected because the measurement uncertainties should have a greater impact on the alpha beam, which is generally less dense than the core and located at higher energy levels. However, it is possible that the larger value of this type of velocity drift be less influenced by transverse velocity fluctuations as it might happen for the proton beam velocity drift.
On the other hand, what we have shown could also indicate that while the proton beam and the alpha beam have a common acceleration mechanism, due to local physical conditions in the plasma, the alpha core shows some discrepancies that can be traced back to the corona's base, where alphas are believed to have undergone acceleration    \citep{asbridge1974,johnson2023}.}

\textcolor{black}{Figure~\ref{Fig10} presents further insights, not disclosed in Figure~\ref{Fig09}, regarding the drift velocities of the proton beam and alpha beam relative to the proton core, specifically focusing on their alignment with the magnetic field. Here we show the distribution of polar (panel A) and azimuthal (panel B) angles in RTN reference system of the orientation of the magnetic field, of the proton beam drift velocity, and of the alpha beam drift velocity with respect to the proton core. 
The distribution confirms a remarkable alignment of both drift velocities to the magnetic field direction. 
Figure~\ref{Fig11} shows, in the same format as the previous Figure, the distributions of polar and azimuthal angles of the magnetic field orientation, the proton beam drift velocity, like in the previous Figure, and the alpha core drift velocity. Compared to the proton beam drift velocity and the alpha beam drift velocity with respect to the proton core, there is a clear lower level of alignment of the alpha core drift velocity with the local magnetic field. Thus, the alpha core drift velocity is confirmed to be the least aligned parameter to the local magnetic field. Larger deviations are shown for larger values of polar and azimuthal angles. 
These observations suggest that the poor focus and long-tail distribution of the alpha beam velocity drift with respect to the alpha core, as depicted in Figure~\ref{Fig09}, is caused solely by misalignment in the velocity drift of the alpha core with respect to the local magnetic field. Obviously, we are left with the problem of understanding the reason for this misalignment, which, as underlined earlier, might also be due to instrumental effects,  but this investigation will be deferred to a future paper.}



\begin{figure*}
    \centering
       \includegraphics[width=18cm]{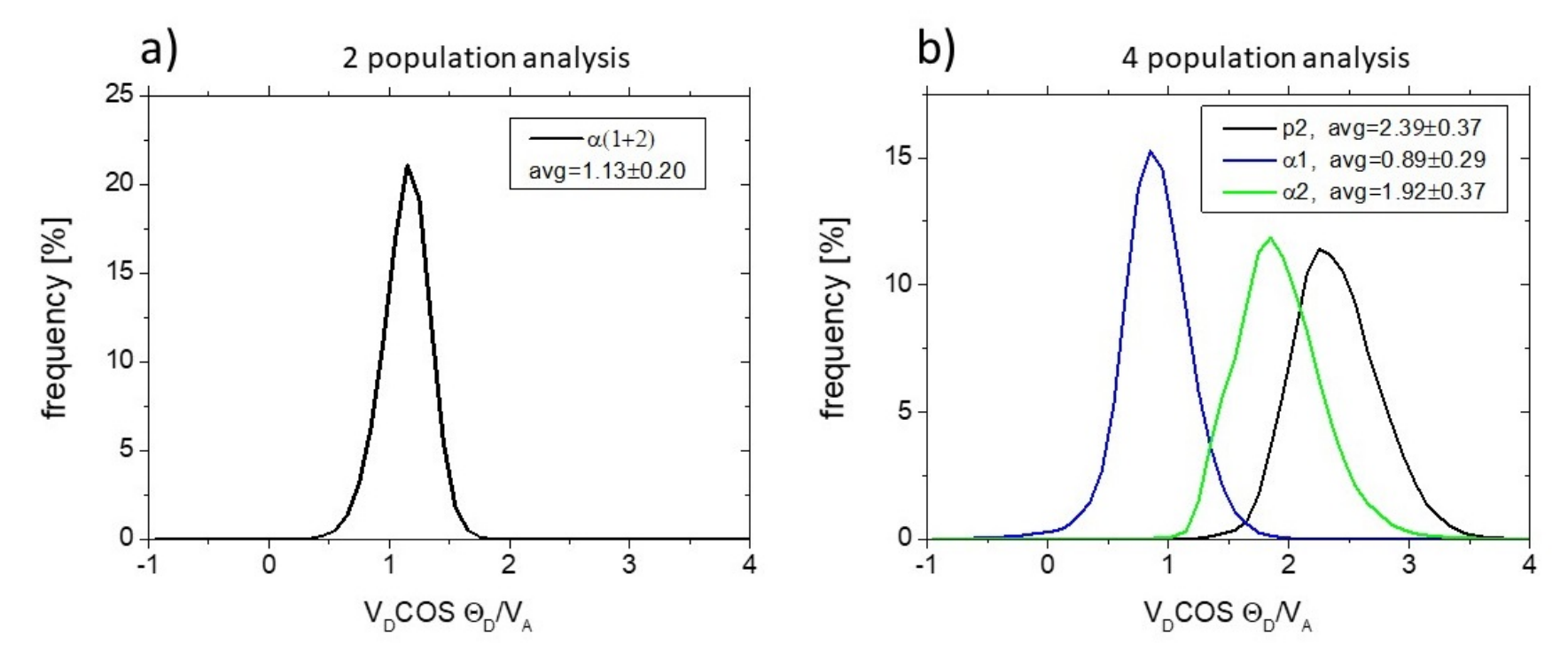}
   \caption{Panel a): frequency histogram of the drift velocity of the alpha population with respect to the proton population as obtained from the 2-population  analysis, projected onto the local magnetic field direction and normalized to the local Alfv\'en speed.\break
   Panel b): frequency histograms of the drift speed for the proton beam and alpha core with respect to the proton core, and alpha beam with respect to the alpha core, as obtained from the 4-population  analysis, projected onto the local magnetic field direction and normalized to the local Alfv\'en speed.} 
    \label{Fig12}
\end{figure*}

Panel a) of Figure \ref{Fig12} shows the distribution of the drift speed $V_D$ of the alpha population with respect to the protons projected onto the local magnetic field direction and normalized by the local Alfv\'en speed $V_A$ as obtained from our 2-population  analysis. The distribution is quite symmetric around the average value of $1.13\pm0.20$. This value is slightly larger than the values reported in the literature which shows that, in general, the drift velocity is of the order of the Alfv\'en velocity that, in our case, has an average value of $65.2\pm 15.1$ km/s \citep{neugebauer1976,reisenfeld2001,marsch1982a,durovcova2017}. 
 Panel b) of Figure \ref{Fig12} shows the distributions relative to the drift speed for the proton beam and alpha core with respect to the proton core, and alpha beam, with respect to the alpha core, projected onto the local magnetic field direction and normalized to the local Alfv\'en speed.
It is known from the literature that  the proton beam drifts at a speed quite larger than the local Alfv\'en speed \citep{marschlivi1987} and that the drift value changes depending on the wind type and the heliocentric distance \citep{marsch1982b}. In our case, the peak of the distribution is around $2.25$ and the average value is $2.39\pm0.37$.
The alpha beam behaves similarly to the proton beam, although it has a lower drift speed whose distribution peaks around $1.85$ Alfv\'en speed and an average value of $1.92\pm 0.37$ Alfv\'en speed. These estimates are new in the literature, since we are not aware of similar estimates made for the alpha core and beam separately.  On the other hand, there are several estimates related to the drift speed of the alpha population as a whole \citep[among others]{marsch1982a,durovcova2017}.

\begin{figure}
    \centering
    \includegraphics[width=12cm]{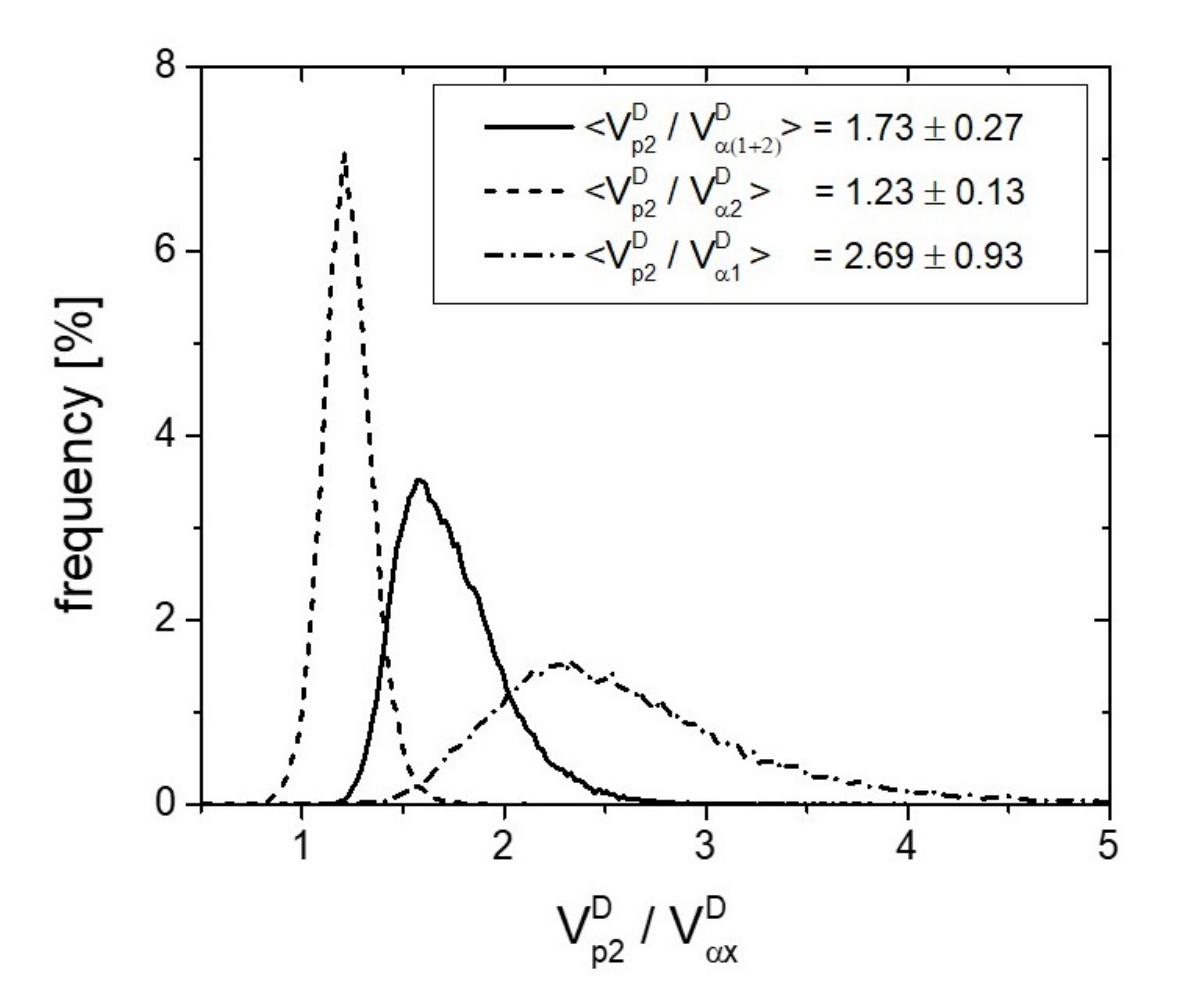}
   \caption{Solid line: distribution of the ratio between the proton beam velocity drift $V^{D}_{p2}$ and that of the entire alpha particle population $V^D_{\alpha (1+2)}$; dashed line: distribution of the ratio between proton beam velocity drift and alpha beam velocity drift $V^D_{\alpha 2}$; dashed-dotted line: distribution of the ratio between proton beam velocity drift and alpha core velocity drift $V_{\alpha 1}^D$.}
    \label{Fig13}
\end{figure}

Following the analysis by \cite{alterman2018}, Figure \ref{Fig13} shows the distribution of the ratio between the velocity drift of the proton beam and that of the total alpha population, as indicated by the solid line. Its average value of $1.73$ remarkably confirms previous estimates reported by \cite{alterman2018} and \cite{demarco2023}. 
It is interesting to note that the analysis performed by \cite{alterman2018} refers to 1 au and covers about $20$ years of data. \cite{demarco2023} analyzed a time interval at $0.65$ au observed in July 2020 while the present analysis refers to data taken by SWA at 0.58 au in September 2022. The only similarity between the data set studied by \cite{alterman2018} and ours is the data selection based on fast and Alfv\'enic wind. 
These results suggest that the value of this ratio ($ \sim 1.7 $) is quite robust and does not change with distance. 
The present analysis, however, offers the possibility to look deeper into this phenomenon allowing us to build the same kind of distribution for the core and the beam of the alpha population. Figure \ref{Fig13} shows these two distributions which appear to be quite peaked for the beam (dashed line) and rather broad for the core (dotted-dashed line). 
Clearly, we expect that the relative density of the alpha core and beam and their drift speeds be all linked together to keep constant the ratio between the proton drift and the alpha drift during the wind expansion, as indicated by \cite{alterman2018}. 
%
%
In addition, we found that the drift speed behaves differently for proton and alpha beams with respect to the alpha core.

\begin{figure*}
    \centering
    \includegraphics[width=18cm]{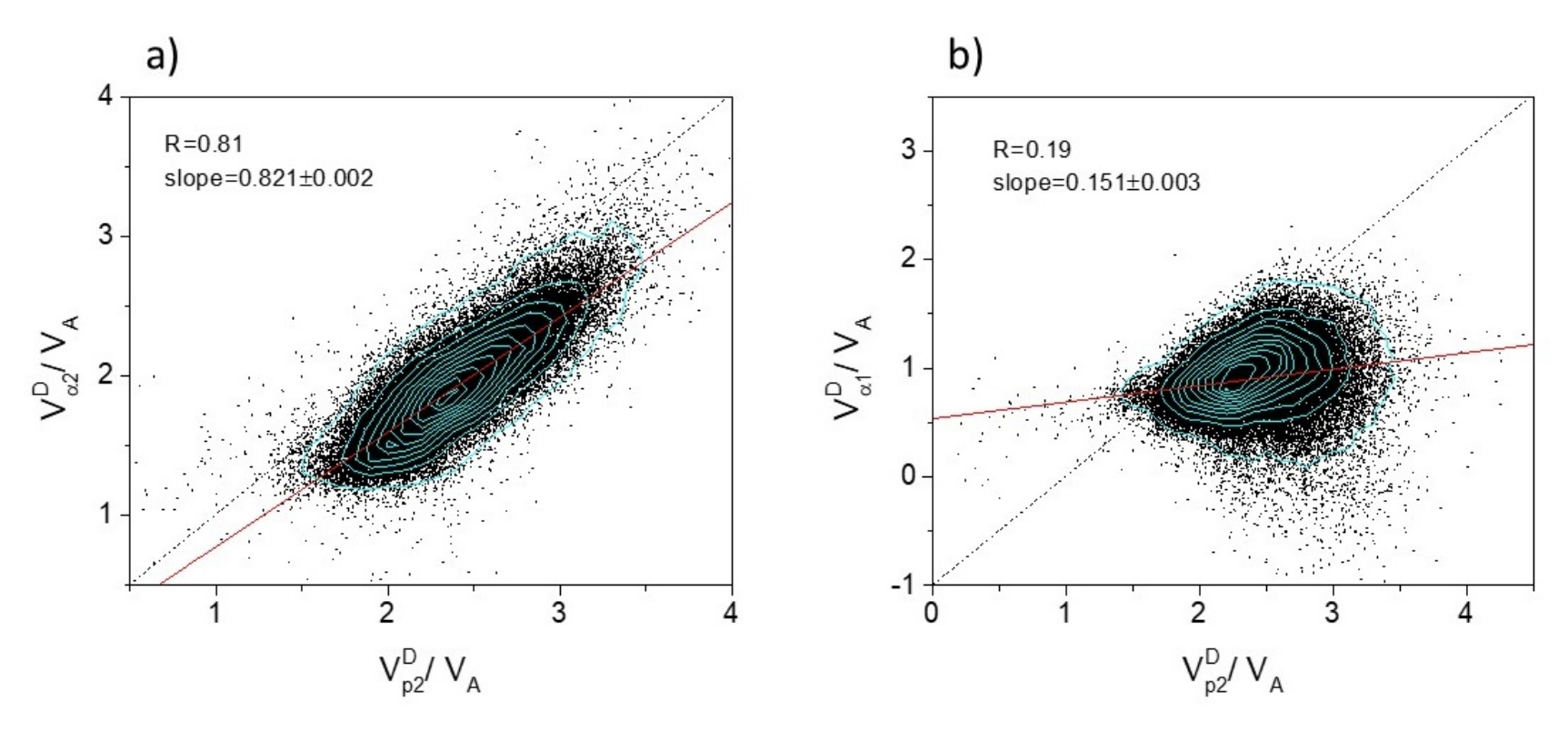}
   \caption{Panel A): scatter plot and contour lines of the velocity drift of the alpha beam ($\alpha$2) vs velocity drift of the proton beam (p2). Panel B): scatter plot and contour lines of the velocity drift of the alpha core ($\alpha$1) vs velocity drift of the proton beam (p2). The dashed line indicates slope=1, and the red line is the linear best fit whose slope is shown in each panel. Contour lines are drawn to guide the eye. }
    \label{Fig14}
\end{figure*}

 Figure \ref{Fig14} displays a significant correlation between the velocity drifts of the proton and alpha beams on panel a, while no meaningful correlation is observed between the velocity drift of the proton beam and the alpha core on panel b. This suggests distinct acceleration mechanisms for the proton and alpha beams, as compared to the alpha core population. As already mentioned above, one possibility is that beams are generated locally by scattering of particles from the core via ion cyclotron waves and or kinetic Alfv\'en waves \citep{marsch1982c,gary2003,araneda2002}. In contrast, the alpha core might have been accelerated at the base of the corona \citep{johnson2023}.

\begin{figure}
   \centering
    \includegraphics[width=13cm]{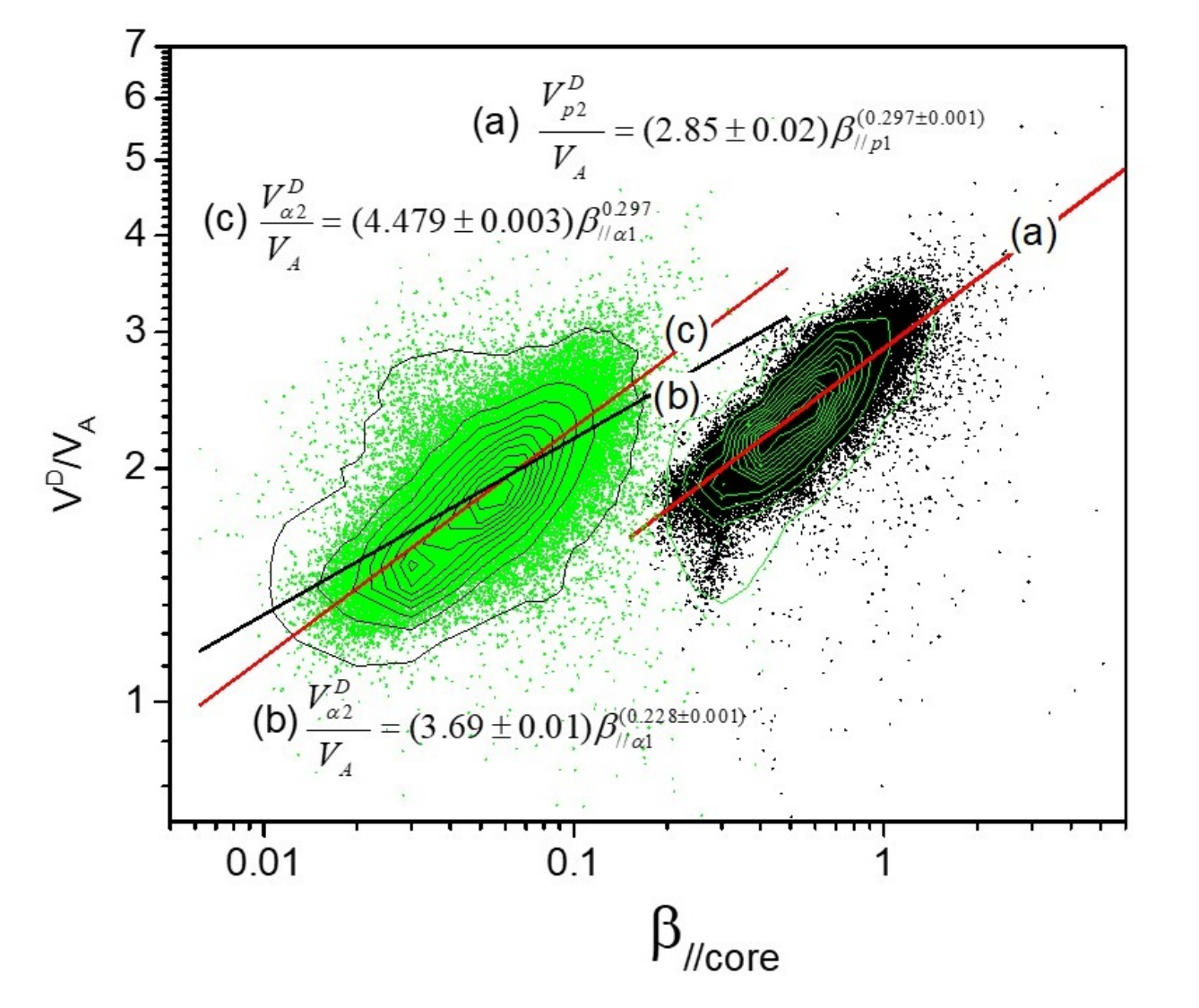}
 \caption{Scatter plots and contour lines of drift speed, \textcolor{black}{normalized to the local Alfv\'en speed,} versus parallel beta for alpha beam (green dots) and proton beam (black dots). The red line (a) is the power law fit to the proton beam population; black line (b) is the power law fit to the alpha beam population; red line (c) is the power law fit to the alpha beam population keeping the same power exponent of curve (a). Contour lines are drawn to guide the eye. 
 }
  \label{Fig15}
\end{figure}

Another feature related to the proton beam drift speed is that this parameter, normalized to the Alfv\'en speed, is related to the proton core plasma beta $\beta_{\|}$ \citep{tu2004,matteini2013}.
Since this relationship constrains the theoretical models describing the generation mechanism of the proton beam in the fast Alfv\'enic wind, we reproduce this study here for both proton and alpha beams. In Figure \ref{Fig15}, we show scatter plots of the normalized velocity drift versus plasma $\beta_{\|}$ for the proton beam population (black dots) and for the alpha beam population (green dots).
%
%
The ratio $V_{p2}^D/V_A$ for the proton beam is plotted versus the parameter $\beta_{\|{core}}=(V_p^{th}/V_A)^2$ where $V_p^{th}$ is the thermal speed for the proton core. Similarly, for the alpha beam, we have used the same $\beta_{\|{core}}$ definition, with the thermal velocity referring to the alpha core. The contour lines on this plane represent the distribution of data points for both populations. The red curve (a) $V_{p2}^D/V_A=(0.285\pm0.002)\beta_{\|{p1}}^{(0.297\pm0.001)}$ is the power law fit to the proton beam population whose expression is not far from \cite{tu2004}, $V_D/V_A=(2.16\pm0.03)\beta_{\|c}^{(0.281\pm0.008)}$, but with a power exponent slightly larger than the one estimated by \cite{demarco2023}  $\sim 0.262$. The black curve (b) $V_{\alpha 2}^D/V_A=(3.69\pm0.01)\beta_{\|{\alpha1}}^{(0.228\pm0.001)}$ is a similar fit applied to the alpha beam population. This curve seems less steep than curve (a) and does not cover completely the central region of the distribution shown by the inner contour lines. This could be because of scattered points, as seen from the bulge on the first contour line. On the other hand, a fitting curve with the same exponent as curve (a) would fit better with all main inner contours of this distribution. 
Thus the proton and alpha beam distributions are more similar than what the power exponent difference shows. 
\textcolor{black}{This would suggest that both distributions follow a $\beta_{//} - V_D/V_A$ correlation similar to that empirically found by \citep{tu2004}, and it could be an additional \textcolor{black}{indication of a common}  beam generation mechanism acting on both protons and alphas. 
It is worth mentioning that \cite{matteini2013} performed a similar study on Ulysses data selected during the first Ulysses north polar transit in 1995–1996, covering the radial distance from 1.3 to about 5 AU and spanning between 30 and 80 degrees heliographic latitude. These authors find some evidence that the marginal stability of the magnetosonic ion beam instability may constrain the data distribution . However, our distributions do not appear to have features that could recall the presence of an upper limit as that observed by \cite{matteini2013}.}

\section{Characterizing Alfvénic fluctuations}
\label{alfenic}



\begin{figure}
    \centering
  \includegraphics[width=11cm]{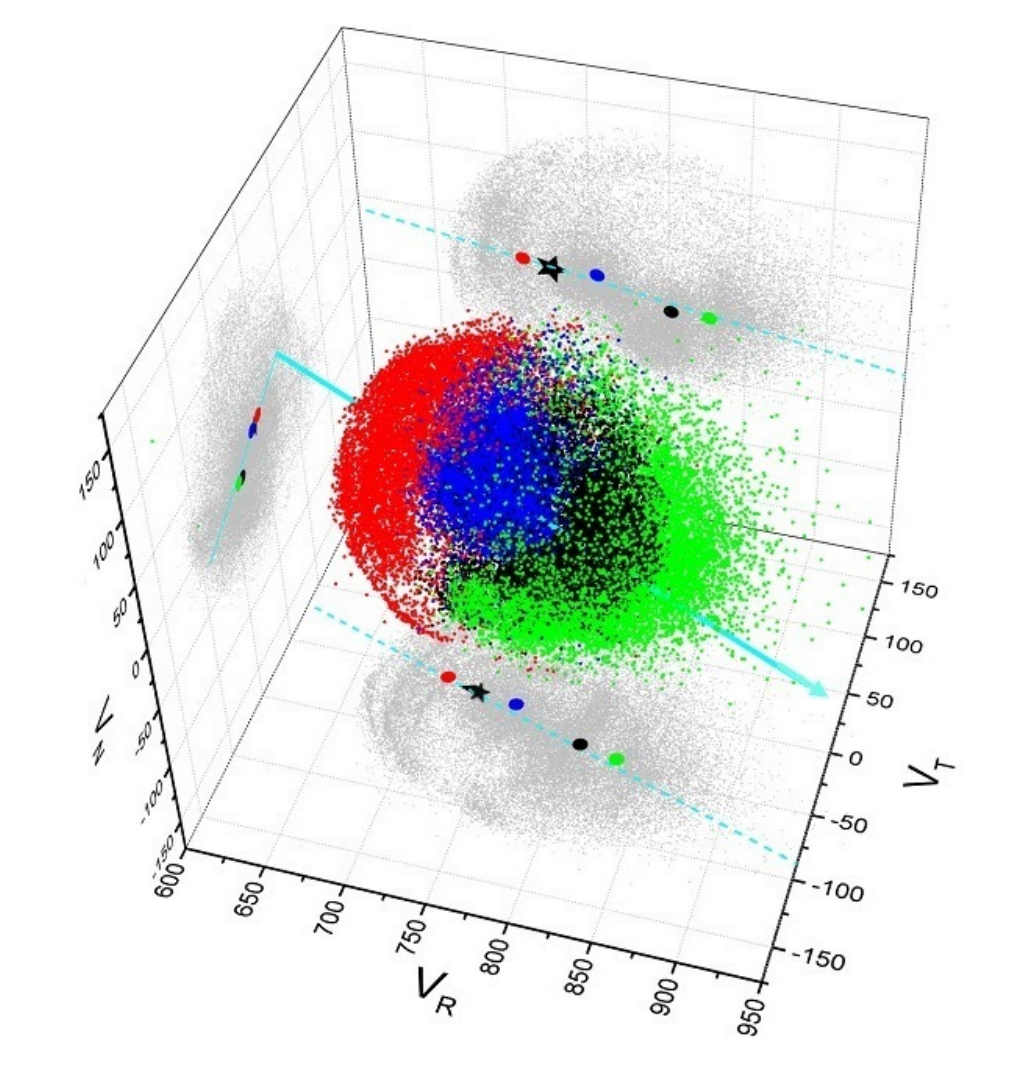}
   \caption{RTN velocity space for proton core (red), proton beam (black), alpha core (blue) alpha beam (green). Projections onto the orthogonal planes are also shown (grey dots). The cyan solid line indicates the average direction of the magnetic field vector during the corresponding time interval. The star symbol on the orthogonal planes shows the projection of the velocity of the center of mass, i.e. the fluid speed, during the whole time interval. Additionally, the colored dots refer to the average speed of each single population. The time interval has been shortened to the first 12 hours of day 16,  for graphic purposes.}
    \label{Fig16}
\end{figure}



In Figure \ref{Fig16}, we show the RTN velocity space for proton core (red), proton beam (black), alpha core (blue) alpha beam (green) populations identified by our code using PAS VDFs at 4 s. The time interval has been shortened to 12 hours, from midnight to noon of day 16,
for graphic purposes. The Figure also shows the projection of each data point onto the orthogonal planes. The cyan solid arrow indicates the average direction of the magnetic field vector during the corresponding time interval. This arrow is also projected onto the orthogonal planes with dots of different colors indicating the location of the bulk velocity for each population. Moreover, the star symbol indicates the projection of the fluid velocity:

\begin{equation}\label{eqn:bulk}
V_{bulk}=\sum\frac{({N_{p1}V_{p1}}+{N_{p2}V_{p2}}+{4N_{\alpha 1} V_{\alpha 1}}+{4N_{\alpha 2}V_{\alpha 2}})}{(N_{p1}+N_{p2}+4N_{\alpha 1}+4N_{\alpha 2)}}
\end{equation}

\textcolor{black}{
It is worth mentioning that in their study, \cite{nemecek2020} suggested that according to the MHD approximation and the dynamics of charged particles in electric and magnetic fields, the velocity of the IMF (de Hoffmann-Teller (HT) velocity) would be considered as the appropriate velocity for solar wind studies. However, the same author concluded that the frame of zero total momentum (fluid frame) and the HT frame could not be the same.}

As expected, the four populations are aligned with the average magnetic field vector. Proton core, alpha beam, and proton beam velocity fluctuations are all distributed over the surface of a sort of hemisphere, as already shown by \cite{bruno2001}, whose concavity for alpha and proton beams is opposite to that of the proton core. This feature is less evident for the alpha core population which appears to be more bubble-like shaped. As we will see in the following, these features depend directly on the value of the drift velocity of each population with respect to the bulk velocity of the center of mass.

\begin{figure*}
    \centering
  \includegraphics[width=18cm]{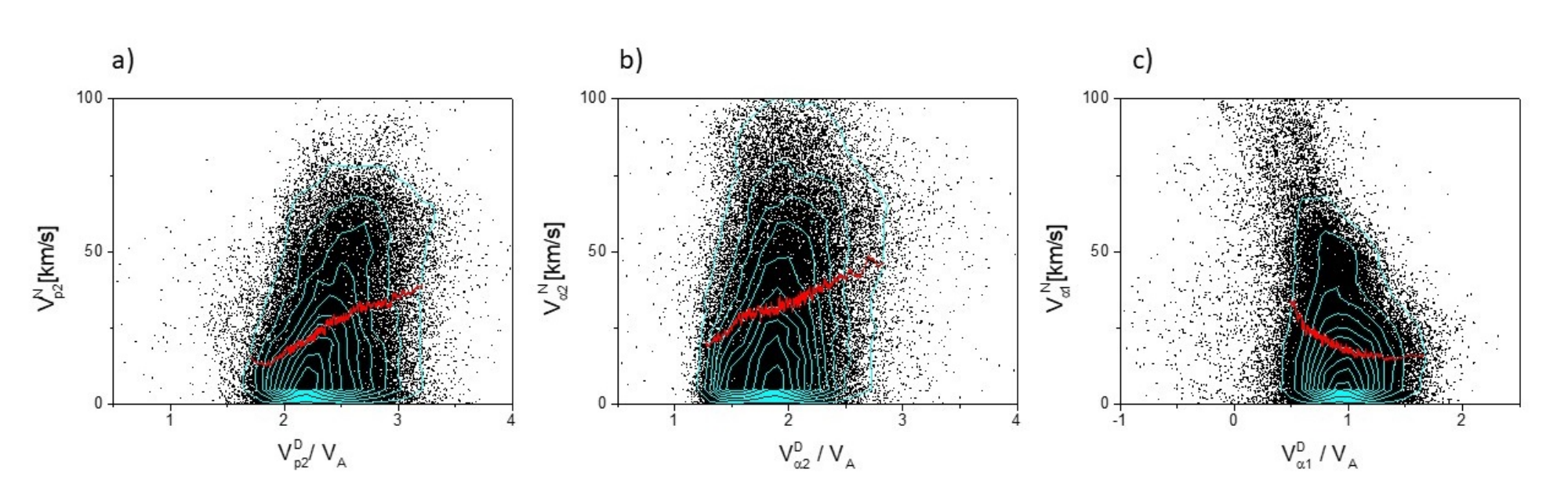}
   \caption{From left to right, histograms of the absolute value of the fluctuations with respect to the mean value of the velocity normal component $V^N$ versus normalized velocity drift for proton beam (p2), alpha beam ($\alpha$2), and alpha core ($\alpha$1), respectively. Contour lines in cyan are also shown in each panel to guide the eye. The red line through each distribution represents the 500-point moving average. }
    \label{Fig17}
\end{figure*}
\begin{figure*}
    \centering
  \includegraphics[width=12cm]{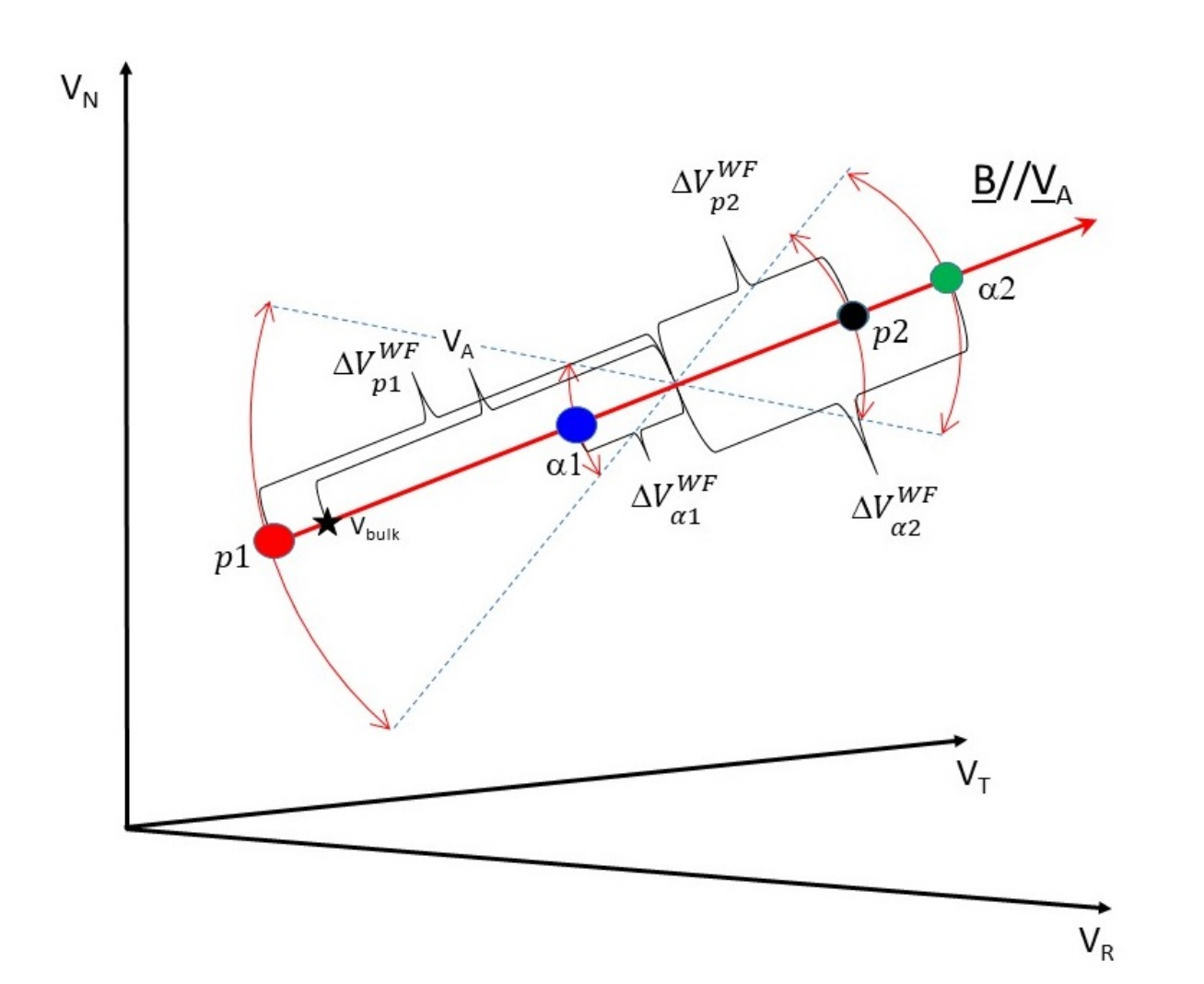}
   \caption{Schematic graphic representation of the phase space in the RTN reference system of proton core (red), alpha core (blue), proton beam (black), and alpha beam (green) velocities fluctuations due to the presence of Alfv\'en waves. The black star symbol represents the estimated fluid velocity calculated from all four populations (see Equation~\ref{eqn:bulk}). The wave frame (also indicated as WF, hereafter) is indicated by the crossing point of the two dashed lines. This point is at $1$ Alfv\'en velocity from the bulk velocity $\vec{V}_{bulk}$. The various $\Delta V^{WF}$ represent the velocity drift of each population from the origin of the WF.} 
    \label{Fig18}
\end{figure*}

Another interesting feature is related to the amplitude of the velocity fluctuations of these different populations. Figure \ref{Fig17} highlights the dependence of the amplitude of velocity fluctuations with respect to their mean value on the drift velocity normalized to the Alfv\'en velocity for the proton beam with respect to the proton core (panel a), alpha beam with respect to the alpha core (panel b) and alpha core with respect to the proton core (panel c). The contour lines of the first two histograms on the left, along with the running average calculated within a window of 500 points (red line), show that the amplitude of both proton and alpha beam velocity fluctuations increases with increasing their respective velocity drift normalized to the local Alfv\'en velocity. 
The opposite behavior is shown by the alpha core population in panel c). Alpha core velocity fluctuations are the largest for values of the normalized velocity drift less than $1$ and decrease as the latter increases. In particular, for normalized velocity drifts $\ge 1$ this dependence fades out. 
The mechanism at the basis of these observations is sketched in Figure \ref{Fig18}. 
As already shown by \cite{matteini2015} and further illustrated by \cite{nemecek2020}, based on early observations by \cite{goldstein1995}, there is a particular frame in phase space that represents the center of the oscillations of both protons and alphas. This frame is the wave frame (also indicated as WF, hereafter). In this frame there is no electric field associated with the fluctuations and each particle's velocity will change only in direction maintaining a constant speed. As a consequence, each particle will be forced to move on the surface of a constant radius in phase space, being the concavity of the resultant hemisphere due to the value of the particle's drift velocity with respect to the wave frame. In particular, each particle will keep its velocity vector aligned to the local magnetic field vector. The wave frame in Figure \ref{Fig18} is represented by the crossing point of the two dashed lines which limit the amplitude of the fluctuations of each population around this pivotal point.
This point is at $1$ Alfv\'en velocity from the bulk velocity, i.e. the velocity of the center of mass, represented by the black star symbol. The Alfv\'en velocity has been computed taking into account all the four different populations, that is:

\begin{equation}\label{eqn:Alfven}
\vec{V}_A=\pm F \frac{\vec{B}}{\sqrt{\textcolor{black}{4\pi} \sum_s\rho_s}} 
\label{alfven}
\end{equation}
where $\vec{B}$ is the IMF vector, $\vec{V}_A$ is the Alfv\'en velocity vector, 
$\rho_s$ is the mass density of the $s^{th}$ species.  The factor $F$ takes into account the thermal and dynamic pressure contribution of protons and alphas \citep[among others]{belcherdavis1971, barnes1979, bavassano1998, alterman2018} and is expressed as follows: 

\begin{equation}\label{eqn:aniso}
  F=\sqrt{1-\frac{4\pi}{|B|^2}\sum_s (P_{s\|}-P_{s\perp}+\rho_s |U_s|^2)} 
\end{equation}
where $P_{\|}, P_{\perp}$ and $U$ are the parallel and perpendicular thermal pressure and the drift velocity with respect to the bulk velocity for each single population indicated by the subscript $s$, respectively.

The different colored symbols in Figure \ref{Fig18} represent the four different populations as identified by the same color code of Figure \ref{Fig16}. The four populations all lie along the local magnetic field direction moving at different speeds with respect to the pivotal point. For the sake of clarity, we list here the meaning of the labels shown in this rather busy figure. The proton core, alpha core,  proton beam, and alpha beam populations are indicated by the labels: $p1$, $\alpha 1$, $p2$, and  $\alpha 2$, respectively. $\vec{V}_A$ is the Alfv\'en velocity, $\vec{B}$ is the local magnetic field, and $\vec{V}_{bulk}$ is the velocity of the center of mass of the whole particle distribution. The parameter $\Delta V^{WF}_{p1}$ is the drift velocity of the proton core from the origin of the wave frame. Similarly,  $\Delta V^{WF}_{\alpha 1}$, $\Delta V^{WF}_{\alpha 2}$ and $\Delta V^{WF}_{p1}$ are the drift velocities of alpha core, alpha beam, and proton beam from the same origin, respectively.
The estimate of $\Delta V^{WF}_{p1}$ is provided by the following expression:
\begin{equation}\label{eqn:DeltaVWF}
\Delta V^{WF}_{p1}=(|\vec{V}_{bulk}|-|\vec{V}_{p1}|)+|\vec{V}_A|
    \end{equation}   
To compute the remaining drifts, we simply substitute $|\vec{V}_{p1}|$ with the velocity corresponding to the other populations.  Consequently, the sign of these drifts will determine whether the velocity fluctuation for a particular sub-population is in phase or anti-phase with respect to the proton core.
These are the correct drift velocities in the wave frame to estimate the amplitude of Alfv\'enic fluctuations associated with different populations. The amplitude of fluctuations increases with larger drift in the wave frame. 
Consequently, the correct amplitude for each of the Alfv\'en velocity components $V^*_{{A}_{i=R,T,N}}$, which correspond to the generic velocity components $V_{i=R,T,N}$ of the generic ion population, should be estimated by the following relation: 
\begin{equation}
\label{eqn:Alfvcomp}
      V^*_{{A}_{i=R,T,N}}=V_{{A}_{i=R,T,N}}\frac{\Delta V^{WF}}{|\vec{V}_{A}|}
\end{equation}

where $V_{{A}_{i=R,T,N}}$  have been obtained from Equation~\ref{eqn:Alfven}, $\Delta V^{WF}$ is the drift velocity of the ion population from the origin of the wave frame, and $|\vec{V}_{A}|$ is the Alfv\'en speed. 



Looking at Figure \ref{Fig18} we should expect that the proton core and alpha core would show an Alfv\'enic correlation of opposite sign with respect to that of the proton beam and alpha beam \citep[see also][]{goldstein1995}. Another interesting inference we derive from the same figure is that 3D Alfv\'enic fluctuations would force these 4 populations to fluctuate on hemispheres of different radii centered around the pivotal point generating the distribution observed in Figure \ref{Fig16}. The case of the alpha core distribution is slightly different. The average drift velocity of this population from the bulk velocity $V_{bulk}$ normalized to $V_{A}$ is $0.78\pm 0.25$ and about $16\%$ of the points have a drift velocity of the order of $V_{A}$. In this situation, most of the velocity distribution, in principle, would be organized on a hemisphere with the same concavity shown by the proton core distribution while the residual part, having a velocity drift of the order of $1$ Alfv\'en velocity would not feel the influence of these waves and would not show any clear concavity. This condition would make the central part of the distribution appear as a bubble as shown in Figure \ref{Fig16}. 



\begin{figure*}
    \centering
  \includegraphics[width=16cm]{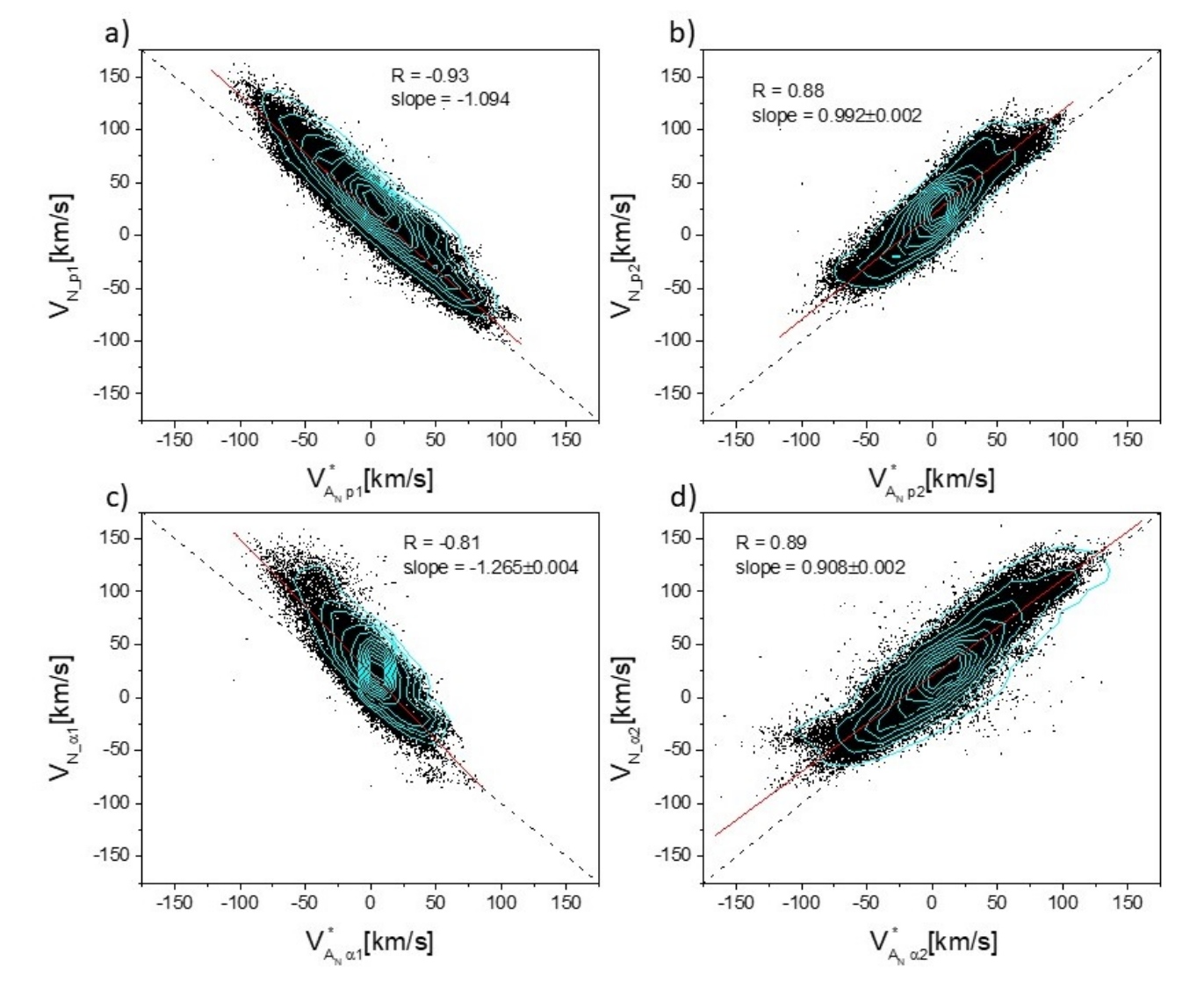}
   \caption{Panels a) to d) refer to proton core (p1), proton beam (p2), alpha core ($\alpha$1), and alpha beam ($\alpha$2), respectively. The horizontal axis for each panel/population shows values of the normal component of velocity fluctuations obtained from Equation~\ref{eqn:Alfvcomp} relative to that population. The vertical axis for each population shows velocity values for the normal component as produced by our code. The Pearson correlation coefficient, R, along with the slope of the fit, represented by the solid red line, is shown in each panel. The dashed line, $V_{N\_ij} = V^*_{{A_N} ij}$ or $V_{N\_ij} = -V^*_{{A_N} ij}$, where i = $p$ or $\alpha$ and j = $1$ or $2$  is shown for comparison. Contour lines are drawn to guide the eye.}
    \label{Fig19}
\end{figure*}

To characterize the Alfv\'enicity of these fluctuations we compare the velocity fluctuations 
with the corresponding magnetic field fluctuations, in Alfv\'en units, obtained from Equation~\ref{eqn:Alfven} for each of the four populations.
 In order to select mainly the Alfv\'enic part of the fluctuations, we evaluate Equation~\ref{eqn:Alfvcomp} only in the $NT$ plane, without considering the radial component $R$, which is the least Alfv\'enic one \citep{brunocarbone2013}, 
%
%
%
and in the following, we show scatter plots relative to the normal component only, which is 
the most Alfv\'enic one \citep{brunocarbone2013}. The four panels of Figure \ref{Fig19}, one for each population, show, on the horizontal axis, values of Alfv\'enic fluctuations obtained from Equation \ref{eqn:Alfvcomp} and on the vertical axis velocity fluctuations. 
All four populations show a rather good level of correlation and, as expected from the schematic graphic representation shown in Figure \ref{Fig18}, proton and alpha cores have an opposite sign of correlation compared to proton and alpha beams. 
Another interesting aspect to notice in Figure \ref{Fig19} is the fact that the absolute value of the slope of the distributions, larger than 1 for the proton and alpha cores and smaller than 1 for the two beams, suggests a slight excess of kinetic energy for proton and alpha cores and vice-versa for proton and alpha beams. In particular, the contour lines in panel C) thicken towards the center and seem to be oriented along the vertical axis highlighting the lack of correlation which characterizes the velocity fluctuations of alpha particles drifting at a speed close to the Alfv\'en speed.
We would like to point out that the level of correlation shown by the alpha populations is particularly striking. The present literature reports that alpha particles, taken as a single population, i.e. core+beam, are not very responsive to Alfv\'enic fluctuations since their velocity drift, so close to the value of the Alfv\'en velocity, makes them surf on the waves with the consequence that they are not influenced by any Alfv\'enic fluctuation \citep{thieme1989,gary2001,matteini2015}. However, this is not always the case if we consider early observations by \cite{goldstein1995}. On the other hand, we showed that the beam is a quite relevant fraction of the whole population and that its velocity fluctuations have an opposite Alfv\'enic correlation compared to the core. These two features are enough to cancel any velocity correlation in case the two populations are not separated.
%
%
%
\begin{figure}
    \centering
  \includegraphics[width=14cm]{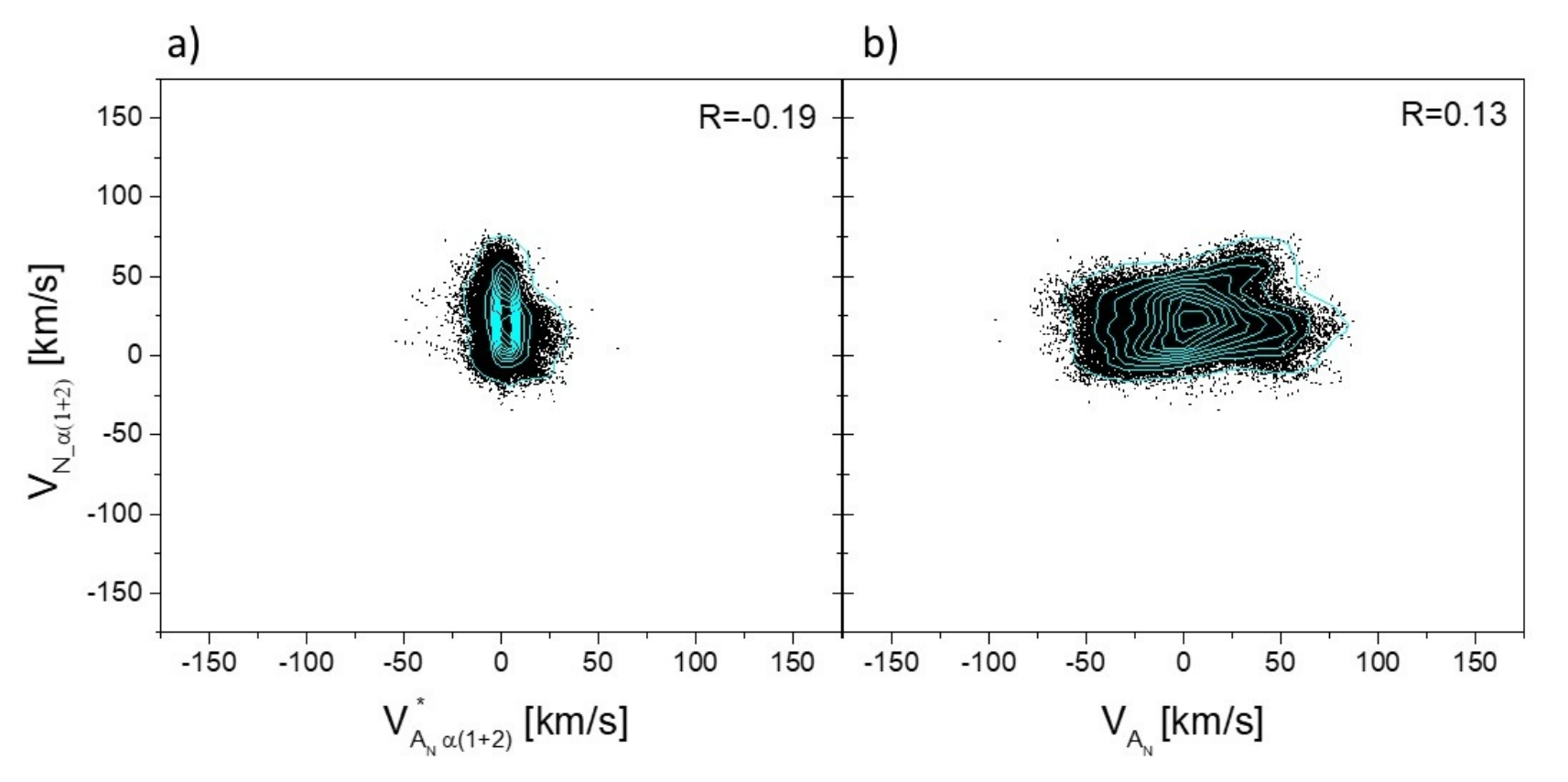}
   \caption{Panel a): Distribution of the normal component of the alpha bulk velocity fluctuations (core and beam not separated) versus the normal component of the Alfv\'en velocity fluctuations obtained from Equation~\ref{eqn:Alfvcomp};
   Panel b): differently from panel a) the normal component of the Alfv\'en velocity fluctuations is obtained from Equation~\ref{alfven}. Contour lines are drawn to guide the eye. }
    \label{Fig20}
\end{figure}
This can be observed in Figure \ref{Fig20}a that 
shows the scatter plot of the velocity fluctuations of the normal component of the bulk velocity of the whole alpha population versus Alfv\'en velocity fluctuations obtained from Equation~\ref{eqn:Alfvcomp}. The amplitude of the fluctuations is enormously reduced and there is a complete lack of Alfv\'enic correlation well at odds with the results shown in the previous Figure \ref{Fig19}, panels c) and d). Similar results (Figure \ref{Fig20}b) are obtained adopting Alfv\'en velocity fluctuations from  Equation~\ref{alfven}, as usually done in the literature.


This remarkable difference with respect to the protons in the presence of Alfv\'enic fluctuations is mainly due to the fact that for the alphas the beam represents a much larger fraction of the whole population compared to the protons and to the fact that the alphas drift from the protons at a speed that is a fraction of the Alfv\'en speed.

\begin{figure*}
    \centering
  \includegraphics[width=16cm]{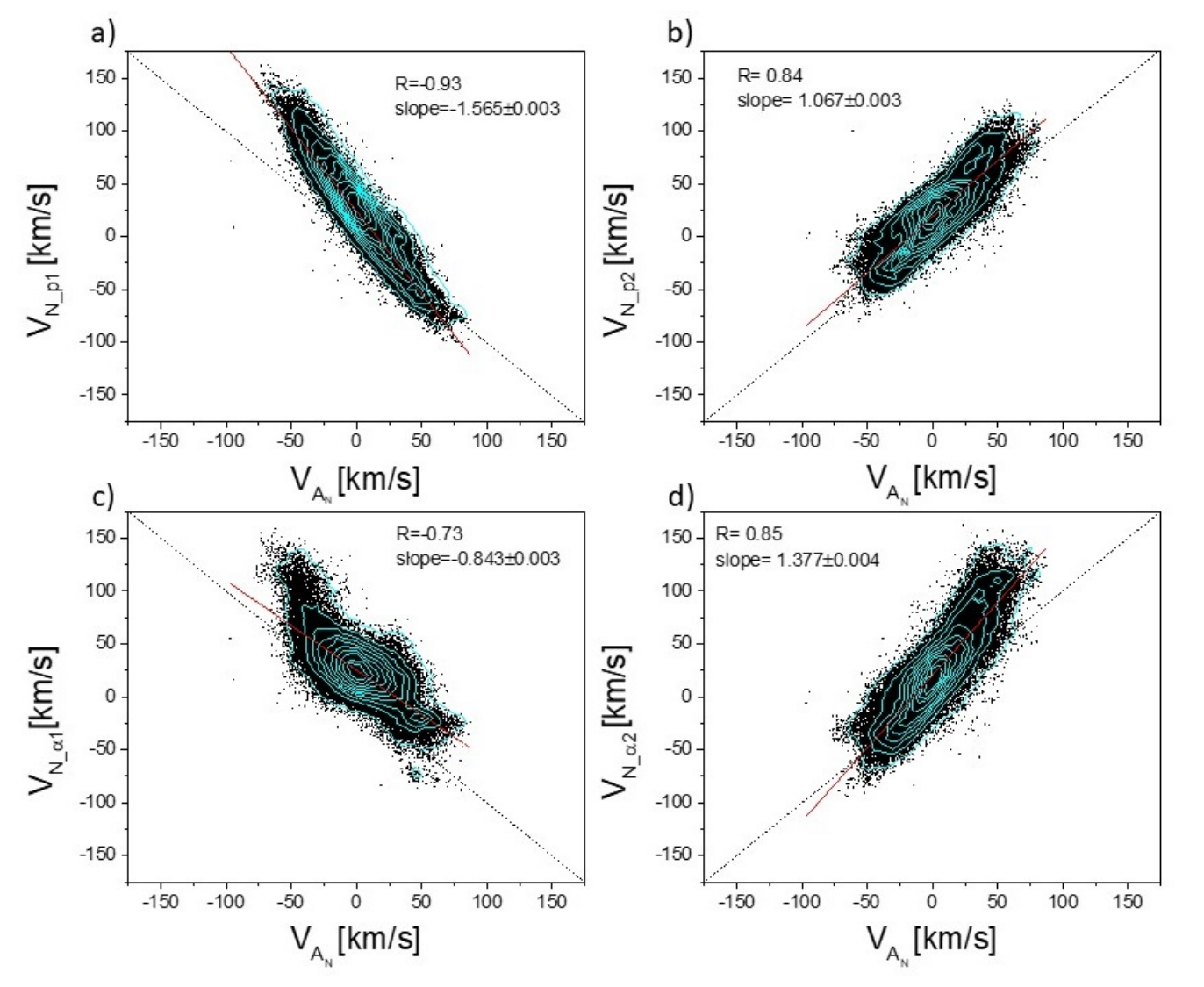}
   \caption{Panels a) to d) refer to proton core, proton beam, alpha core, and alpha beam, respectively. The horizontal axis for each panel/population shows values of the normal component of velocity fluctuations obtained from Equation~\ref{alfven}. The vertical axis for each population shows velocity values for the normal component as produced by our code. The Pearson correlation coefficient, R, along with the slope of the fit, represented by the solid red line, is shown in each panel. The dashed line, $V_{N\_ij} = V_{{A_N}}$ or $V_{N\_ij} = -V_{{A_N}}$, where i = $p$ or $\alpha$ and j = $1$ or $2$  is shown for comparison. Contour lines are drawn to guide the eye.}
    \label{Fig21}
\end{figure*}

\textcolor{black}{In the current context, it is intriguing to visualize the distributions, which are displayed in Figure \ref{Fig19}, using the amplitude of the Alfv\'enic fluctuation derived from Equation \ref{alfven}, as is typically done in the literature, instead of scaling it through Equation \ref{eqn:Alfvcomp}. Figure \ref{Fig21} illustrates this comparison, where we present the values of the normal component of the Alfv\'en velocity computed from Equation \ref{alfven} on the X-axis, using the same format as Figure \ref{Fig19}. We observe a general deterioration in the degree of correlation, except for the proton core distribution, as expected. Specifically, the alpha core distribution (panel c) is significantly distorted for high velocity values. Additionally, both the proton core and alpha beam distributions deviate significantly from equipartition. This comparison reinforces the validity of our analysis, which is based on the mechanism depicted in Figure \ref{Fig18}.}

\section{Considerations about the alpha beam population}
\label{alpha}

Being an electrostatic analyzer, SWA-PAS detects the ions by measuring their kinetic energy $E$ per unit charge $q$, therefore there is the possibility that the population we assume to be the beam of the alpha particles in reality might be a population of oxygen ions O$^{6+}$.  These ions, roughly traveling at the same speed as the protons, would appear to have an $E/q\sim 2.67$ being their mass equal to $16$ proton masses and their charge equal to $6$ proton electric charges. As a consequence, the expected bulk speed of these ions would appear to be $\sqrt{2.67}=1.63$ times higher than the proton core bulk speed, which in our case is $\sim$ 700 km/s. If we also add a velocity drift around $80\%$ (just to be consistent with the drift we observe for the core of the alphas $\sim$ 75 km/s) of the Alfv\'en speed \citep{marsch1982a}, this value would be around $1.72$ times larger than the proton core speed. On the other hand, the ion population that we identify as a beam of alpha particles has an average speed of $\sim 1.19$ times faster than the proton core and, being helium ions, SWA-PAS detect this population at a speed $\sqrt{2}\times 1.19\sim 1.68$ times faster than the proton core speed. Thus, there is a chance to confuse what we identify as an alpha beam with a population of O$^{6+}$. However, there is  another parameter that we can consider to strengthen our original assumption: the relative density. As we already saw in Figure \ref{Fig04} the relative density of the beam with respect to the alpha core peaked at a value slightly less than $50\%$. 
On the other hand, there are consistent observations in the literature about the relative abundance O/He. Just to cite a few examples, \cite{bochsler1986} found that oxygen ions are about $1.3\%$ of helium ions in the solar wind, in good agreement with a successive determination by \cite{collier1996} who estimated this ratio to be around $1.4\%$. Both values do differ enough from the value of $\sim 50\%$ represented by the population that we identify as an alpha beam.
In conclusion, our analysis confirms previous inferences by \cite{feldman1974,marsch1982a} who noticed non-thermal double-peaked spectra of solar wind helium distributions observed at 1 au and within the inner heliosphere, respectively.


\section{Summary and Conclusions}
\label{conclusioni}
This paper discusses the observations of solar wind plasma made by the PAS sensor, which is one of the four sensors in the plasma suite SWA onboard the Solar Orbiter spacecraft. During September 2022, the spacecraft observed a high-speed stream while it was $0.58$ au away from the Sun.  We have chosen the region of the stream with the highest Alfv\'enicity, lasting about 3 days, because a high degree of Alfv\'enicity is an important condition that may be closely related to the occurrence of non-Maxwellian features, like temperature anisotropies and proton and alpha beams. This proved to be a rewarding choice as our code localized proton and alpha beams for approximately $90\%$ of all VDFs recorded by PAS.  After this first step, about 56000 VDFs were analyzed to characterize and compare the kinetic features of the core and beam populations of protons and alpha particles.
Our analysis estimated an alpha content of about $4\%$ of the proton population, well in agreement with numerous previous estimates reported in literature \citep[among others]{bochsler2007, feldman1978, borrini1982, marsch1982a}{}. 
We demonstrated the existence of a beam for alpha particles that cannot be mistaken for the O$^{6+}$ population. However, we cannot rule out the possibility of contamination by O$^{6+}$. Interestingly enough, we found that the alpha beam, differently from the proton beam, represents a relevant fraction, about $\sim 50\%$, of the core population. In this respect, the proton beam, within the analyzed time interval, is much less relevant being only about $10\%$ of the proton core, in good agreement with previous studies based on Helios fast wind observations \citep{marsch1982a,marsch1982b,durovcova2019}.  
Such a large alpha beam plays an important role in estimating the kinetic features of the alphas when the core and beam are not analyzed separately.  For instance, we showed that the invoked additional anomalous overheating mechanism \citep{kasper2008} necessary to justify a total temperature ratio between alphas and protons $\sim 5$ is not needed if we restrict this ratio to the core only.  As a matter of fact, in this case,  we found a total temperature ratio $\sim 4$ that suggests an equal thermal speed for the core populations of protons and alphas  \citep{kasper2008}. Obviously, this does not solve the overheating problem that still requires a physical explanation, for instance in terms of Alfv\'en-cyclotron dissipation mechanism, which favors heavier ions with respect to protons \citep{kasper2017,kasper2008,ofman2002,feldman1974,neugebauer1976, neugebauer1981,feynman1975}.  It is worth noting the striking similarity between the histograms of the total temperature ratio between the beam and core for protons and alphas suggesting a possible shared heating mechanism for both beams.
Our analysis allows us to show and compare the temperature anisotropy of the core and beam not only for the protons but also for the alphas. We found values of anisotropy for protons and alphas around $2$, for all the  core and beam populations. These estimates confirm previous studies of the anisotropy of the proton core and provide new insights into the anisotropy of the proton and alpha beams, but do not appear to agree with the values for the anisotropy of the alpha core, generally $< 1$, found in the literature \citep{marsch1982a, kasper2008}. In particular, \cite{marsch1982b} reported that in the fast wind, the core part of the helium distribution shows some indications for $T_{\|}>T_{\perp}$.
In addition, \cite{kasper2008} reported higher values of $T_{\|}>T_{\perp}$ for helium ions for smaller values of collisional age also identifiable with fast streams. However, in the latter case, we are not aware of any specific separation between the alpha core and beam.
The study of the velocity drift of the various populations revealed that the alpha core drifts with respect to the proton core at a velocity less than the local $V_A$, the proton beam with respect to the proton core, and the alpha beam with respect to the alpha core both move at a velocity of about twice $V_A$. This latter consideration might suggest that the two beams might have been accelerated by a different mechanism than the one that acted on the alpha core. Present literature indicates that while beams are generated locally by scattering of particles from the core by ion cyclotron waves and or kinetic Alfv\'en waves, the alpha core is accelerated directly at the base of the corona \citep[][and references therein]{johnson2023}
To strengthen the suggestion in favor of a local generation of proton and alpha beams there is also the fact that the distribution of the alpha beam drift angle is remarkably similar to that of the proton beam when the drift velocity vector 
is estimated with respect to the proton core. On the other hand, the distribution of the drift velocity vector relative to the alpha core is as broad as that of the alpha core relative to the proton core. The remarkable width of the distribution of the drift angle for the alpha core can not depend on the weaker statistics associated with the alphas otherwise this problem would affect even more the same kind of distribution for the alpha beam. We support the idea of a non-local acceleration of the alpha core which might reside instead at the base of the corona.
We found also other observations in favor of a shared generation mechanism for the proton and alpha beams like the remarkable correlation between their drift velocities. This is particularly relevant if compared to the total lack of correlation between the drift velocities of the core and alpha beam. The two beam populations behave in a similar way also regarding the empirical law found by \cite{tu2004} relating the beam velocity drift, normalized to the local Alfv\'en speed, and the $\beta_{\|}$ relative to the core of the respective population.

Finally, we analyzed the Alfv\'enicity of all 4 ion families within the wave frame WF as identified by the wave propagation speed \citep{goldstein1995,matteini2015}. 
In this particular frame in phase space, which represents the center of the oscillations of protons and alphas, the waves are at rest.
We tested the Alfv\'en relation for each population considering the velocity drift $\Delta V^{WF}$ from this pivotal point of the oscillations (see Equation~\ref{eqn:Alfvcomp}) and found a rather good Alfv\'enic correlation for all four populations.
As expected \citep{goldstein1995}, the sign of the correlation of the beams is opposite to that of their respective cores, depending on the value of $\Delta V^{WF}$. On the other hand, we found no correlation at all and remarkably small amplitude fluctuations when the alphas were treated as a single population, confirming earlier findings \citep{thieme1989, goldstein1996, matteini2015}. This is due to the fact that the alpha beam is rather massive and its velocity fluctuations have an opposite sign compared to the alpha core. Consequently, the center of mass of this core-beam system appears to be not sensitive to the presence of Alfv\'enic fluctuation.

We are rather confident that the kind of analysis shown in this paper, based on a careful examination of the kinetic features of core and beam separately for proton and alphas, is the correct way to understand better the physical processes that shape the particle distribution in the solar wind.

\begin{acknowledgements}
This work is supported by the grant “Machine Learning on Solar Wind Velocity Distribution Functions” financed by the National Institute for Astrophysics under the call “Fundamental Research 2023”. Solar Orbiter is a space mission of international collaboration between ESA and NASA, operated by ESA. Solar Orbiter Solar Wind Analyser (SWA) data are derived from scientific sensors which have been designed and created, and are operated under funding provided in numerous contracts from the UK Space Agency (UKSA), the UK Science and Technology Facilities Council (STFC), the Agenzia Spaziale Italiana (ASI), the Centre National d’Etudes Spatiales (CNES, France), the Centre National de la Recherche Scientifique (CNRS, France), the Czech contribution to the ESA PRODEX programme and NASA. Solar Orbiter SWA work at INAF/IAPS is currently funded under ASI grant 2018-30-HH.1-2022.
The authors acknowledge T. Horbury and the MAG Team for Solar Orbiter magnetic field data. The authors warmly thank the Referee for stimulating comments and suggestions.


\end{acknowledgements}

\end{document}